\newcommand{\paperOption}{compsoc}
\newcommand{\usehyperref}{\usepackage[hidelinks]{hyperref}}

\InputIfFileExists{option1}{%
    \typeout{Options overridden!}%
}{}

\documentclass[\paperOption]{IEEEtran}
\usepackage{amsmath,amssymb}
\usepackage{mathtools}
\usepackage{graphicx}
\usepackage[noadjust]{cite}
\usepackage{algorithm,algorithmicx,algpseudocode}
\usepackage{booktabs}
\usehyperref

\newtheorem{theorem}{Theorem}
\newtheorem{corollary}{Corollary}
\newtheorem{lemma}{Lemma}
\newtheorem{proposition}{Proposition}
\newtheorem{remark}{Remark}
\newtheorem{definition}{Definition}

\newtheorem{problem}{Problem}

\newcommand{\eqdef}{:=}

\newcommand{\relvar}[2]{\buildrel \text{#1} \over #2}
\newcommand{\eqvar}[1]{\relvar{#1}{=}}
\newcommand{\geqdef}{\triangleq}

\newcommand{\gevar}[1]{\relvar{#1}{\ge}}
\DeclarePairedDelimiterX{\clip}[3]{\langle}{\rangle_{#2,#3}}{#1}
\DeclarePairedDelimiterX{\lowerclip}[2]{\langle}{\rangle_{\ge #2}}{#1}
\DeclarePairedDelimiterX{\upperclip}[2]{\langle}{\rangle_{\le #2}}{#1}
\newcommand{\term}[1]{\emph{#1}}

\newcommand{\integers}{\mathbb{Z}}
\newcommand{\integersOf}[1]{\integers_{#1}}
\newcommand{\pintegers}{\integersOf{>0}}
\newcommand{\nnintegers}{\integersOf{\ge 0}}
\newcommand{\real}{\mathbb{R}}
\newcommand{\nnreal}{\real_{\ge 0}}
\newcommand{\preal}{\real_{> 0}}

\newcommand{\expect}{\mathbb{E}}
\newcommand{\var}{\mathrm{Var}}
\newcommand{\cnu}{\pi^{\mathrm{cn}}}

\newenvironment{sketch}{\begin{IEEEproof}[Sketch of Proof]}{\end{IEEEproof}}
\newcommand{\proofref}[2][Proof]{\hfill\hbox{(#1 #2)}}

\begin{document}
    \title{Low-Complexity AoI-Optimal Status Update Control with Partial Battery State Information in Energy Harvesting IoT Networks}

    \author{
        Hao~Wu,
        Shengtian~Yang,~\IEEEmembership{Senior~Member,~IEEE},
        Jun~Chen,~\IEEEmembership{Senior~Member,~IEEE},
        Chao~Chen, Anding~Wang%
        \thanks{Corresponding author: Shengtian Yang.}
        \thanks{H.~Wu, S.~Yang, C.~Chen, and A.~Wang are with the School of Information and Electronic Engineering (Sussex Artificial Intelligence Institute), Zhejiang Gongshang University, Hangzhou 310018, China (e-mail: \mbox{wu\_hao\_a@126.com}, \mbox{yangst@codlab.net}, \mbox{eckio\_491@zjgsu.edu.cn}, \mbox{anding\_704@hotmail.com}).}%
        \thanks{J.~Chen is with the Department of Electrical and Computer Engineering, McMaster University, Hamilton, ON L8S 4K1, Canada (e-mail: chenjun@mcmaster.ca).}%
    }

    \maketitle

    \begin{abstract}
        For a two-hop IoT system consisting of multiple energy harvesting sensors, a cache-enabled edge node, and multiple monitors, the status update control at the edge node, which has partial battery state information (pBSI) of the sensors, is formulated as a pBSI problem.
        The concept of inferred pBSI is introduced to reduce the noiseless single-sensor pBSI problem to a Markov decision process with a moderate state-space size, enabling the optimal policy to be obtained through a value iteration algorithm.
        A lower bound on the expected time-average on-demand age of information performance is established for the general single-sensor status update problem.
        For the single-sensor pBSI problem, a semi-closed-form policy called the current-next (CN) policy is proposed, along with an efficient post-update value iteration algorithm with a per-iteration time complexity proportional to the square of the battery capacity.
        A weighted-update-gain-competition (WUGC) approach is further leveraged to extend the CN policy to the multi-sensor case.
        Numerical results in the single-sensor case demonstrate the near-optimal performance of the CN policy across various energy arrival processes.
        Simulations for an IoT system with $100$ sensors reveal that the WUGC-CN policy outperforms the maximum-age-first policy and the random-scheduling-based CN policy under Bernoulli energy arrival processes.
    \end{abstract}

    \begin{IEEEkeywords}
        Age of information, energy harvesting, low complexity, partial battery state information, status update control.
    \end{IEEEkeywords}

    \section{Introduction}\label{sec-introdcution}

    Status freshness is a vital concern in time-sensitive applications of the Internet of Things (IoT).
    The age of information (AoI) metric, defined as the time elapsed since the generation of the last successfully received status update, provides a promising way to quantify status freshness.
    In energy harvesting (EH) IoT networks, IoT devices are usually powered by batteries with energy harvesters.
    In this context, the status-freshness issue boils down to the design of AoI-optimal status update control, subject to available energy supply (see, e.g.,~\cite{
        kaul2012,
        bacinoglu2015,
        yang2016,
        arafa2017,arafa2017-2,bt2017,wu2017,yates2017,
        arafa2018,bt2018-1,farazi2018,kri2018,
        arafa2019,arafa2019-2,arafa2019-3,bt2019,chu2019,hzhu2019,leng2019,ma2019,tunc2019,yang2019,
        hatami2021,yates2021review,
        hatami2022,hatamipartial,hatami2024status,
        chen2024peak})
    and radio resource constraints (see, e.g.,~\cite{
        kadota2018opti,
        bedewy2021optimal,hatami2021,kadota2021mini,
        chen2022uncertainty,hatami2022,fang2022age}).
    The battery levels of the devices
    are among the crucial pieces of information
    for status update control.
    In terms of the nature of available battery state information (BSI), status update control can be divided into three categories:
    (i) status update control with exact BSI (eBSI), (ii) status update control with partial BSI (pBSI), and (iii) status update control with no BSI (nBSI).

    There is a large body of literature on the design of status update control policies with eBSI  (e.g.,~\cite{arafa2017,farazi2018,bacinoglu2015,arafa2017-2,arafa2019,arafa2018,arafa2019-2,bt2019,leng2019,tunc2019,arafa2019-3,hzhu2019,hatami2022}) and nBSI (e.g.,~\cite{yang2016,wu2017,yang2019} and the references therein).
    In contrast,  only preliminary studies have been conducted on the pBSI case.
    The problem of designing status update control policies with pBSI (hereafter referred to as the pBSI problem) was first introduced in~\cite{hatami2021}.
    In~\cite{hatamipartial} and~\cite{hatami2024status}, a simple case of the pBSI problem, where the channels for status updating are noiseless, was modeled as a partially observable Markov decision process (POMDP); by reformulating it as an equivalent belief-MDP and truncating the associated belief space into a finite space, a dynamic programming algorithm was presented to obtain an approximately optimal policy.
    However, the complexity of this algorithm is remarkably high compared to similar algorithms in the eBSI case.

    On the other hand, designing AoI-optimal status update control in a multi-device IoT network with radio resource constraints is challenging due to the rapidly increasing complexity with the number of sensors.
    Several studies have focused on near-optimal solutions under different system settings but with no energy-supply constraint, such as Whittle’s index policy~\cite{kadota2018opti}, the maximum-age-first (MAF) policy~\cite{bedewy2021optimal}, and the max-weight policy~\cite{kadota2021mini}.
    These schemes, when applied to EH IoT networks, naturally fall into the nBSI case.
    For the eBSI case, low-complexity near-optimal algorithms have also been provided (e.g.,~\cite{hatami2022}).
    For the pBSI scenario, a deep reinforcement learning-based approach was introduced in~\cite{xu2024optimal} to minimize the discounted average age of correlated information.
    However, this method suffers from high computational complexity, and its optimality remains unverified.

    In practice, when the status updating procedure involves multiple devices, exact knowledge of all BSI is often unavailable due to limited energy supply and radio resources.
    In particular, in scenarios where status update control is performed at a separate control node, which is common in IoTs, the knowledge of BSI at the control node is only available via status updates and hence may not be fresh for status update control.
    A pBSI model is evidently more suitable for such cases, and a low-complexity solution to this model is highly anticipated both theoretically and practically.

    The above facts and ideas thus motivate this work, which focuses on the pBSI problem in a two-hop IoT system with a cache-enabled edge node as the control and relay node for status update.
    Our goal is to design a low-complexity status update control scheme with pBSI at the edge node to minimize the average AoI of status received by the monitors in the system.

    The main contributions of this paper are as follows.

    (1) The concept of inferred pBSI is introduced to reduce the noiseless single-sensor pBSI problem to an MDP with a moderate state-space size.
    This enables us to obtain the optimal policy using a value iteration algorithm with a per-iteration time complexity of $\mathrm{O}(\overline{B}\overline{\Delta} (\overline{B}+\overline{D}))$ (Corollary~\ref{co:noiseless_pbsi_vi_complexity}), lower than that of~\cite[Alg.~2]{hatami2024status} for solving the noiseless pBSI problem under Bernoulli energy arrival processes, where $\overline{B}$, $\overline{\Delta}$, and $\overline{D}$ are the battery capacity, the maximum admissible age of cached status information, and the maximum admissible age of failure-inferred battery level, respectively.

    (2) A lower bound on the expected time-average on-demand AoI performance (Theorem~\ref{th:lower_bound}) is established for the general single-sensor status update problem, regardless of whether it involves eBSI, pBSI, or nBSI.

    (3) The concept of inferred pBSI is generalized to address the general single-sensor pBSI problem.
    A semi-closed-form policy, termed the current-next (CN) policy, is proposed.
    This policy relies on estimating the so-called post-update value function, for which we develop an efficient value iteration algorithm using an online-offline hybrid approach.
    The per-iteration time complexity of this algorithm is $\mathrm{O}(\overline{B}^2)$ (Proposition~\ref{pr:smdp_size}), significantly lower than the $\mathrm{O}(\overline{B}^2\overline{\Delta})$ time complexity required for a general value iteration algorithm in the eBSI case with general energy arrival distributions.
    By eliminating the dependence on $\overline{\Delta}$ in the time complexity, the CN policy runs efficiently even for large $\overline{\Delta}$ values, overcoming a limitation of prior value-iteration-based methods such as those in~\cite{hatami2021,gindullina2021ageofinformation,hatami2022}.
    Furthermore, we propose the weighted-update-gain-competition (WUGC) approach to extend the CN policy to the multi-sensor case with a maximum simultaneous-update constraint.
    Numerical results in the single-sensor case demonstrate that the CN policy is near optimal under various energy arrival processes.
    Simulations for an IoT system with $100$ sensors reveal that the WUGC-CN policy outperforms the MAF policy and the random-scheduling-based CN (random-CN) policy under Bernoulli energy arrival processes.

    The rest of this paper is organized as follows. Section~\ref{sec:system-model} presents the IoT system model and formulates the problem with key definitions.
    In Section~\ref{sec:noiseless_pbsi}, we reduce the noiseless single-sensor pBSI problem to an MDP with a moderate state-space size, for which we obtain the optimal policy using a value iteration algorithm.
    Section~\ref{sec:general_pbsi} introduces the CN policy and an efficient post-update value iteration algorithm to solve the general single-sensor pBSI problem.
    Section~\ref{sec:proofs} presents the proofs of the main results in Section~\ref{sec:general_pbsi}.
    The WUGC approach is then presented in Section~\ref{sec:wugc} to extend the CN policy to the multi-sensor case.
    In Section~\ref{sec:simulation}, we analyze the structural properties of the CN policy using numerical methods, followed by an evaluation of the CN and WUGC-CN policies in the single-sensor and multi-sensor cases, respectively.
    Finally, Section~\ref{sec:conclusion} concludes the paper.

    Throughout this paper, the symbol $\geqdef$ is used to denote a global definition, while the symbol $\eqdef$ is used to indicate a definition that is valid only within a local scope (e.g., a section or a proof).
    The set of all integers in a set $A$ is denoted by $\integersOf{A}\geqdef \integers\cap A$, with $\nnintegers$ and $\pintegers$ being shorthand notations for $\integersOf{[0,+\infty)}$ and $\integersOf{(0,+\infty)}$, respectively.
    The clipped value of $x$ in the interval $[a,b]$ is denoted by $\clip{x}{a}{b}$, with $\upperclip{x}{b}$ and $\lowerclip{x}{a}$ serving as shorthand notations for $\clip{x}{-\infty}{b}$ and $\clip{x}{a}{+\infty}$, respectively.

    \section{System Model and Problem Formulation}\label{sec:system-model}

    \subsection{Network Model}\label{subsec:network}

    \begin{figure}[htbp]
        \centering
        \includegraphics{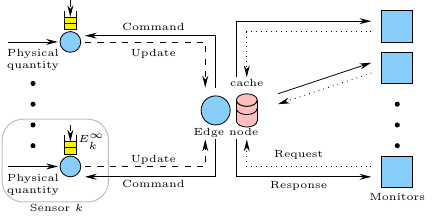}
        \caption{A two-hop IoT network consisting of multiple sensors, one edge node, and multiple monitors.}
        \label{fig-system-model}
    \end{figure}

    \begin{table}[htbp]
        \centering
        \caption{The main system parameters and their symbols}
        \label{tab:para-sym}
        \begin{tabular}{@{}lll@{}}
            \toprule
            & Main system parameter                                            & Symbol                \\ \midrule
            & The number of sensors                                            & $K$                   \\
            & Maximum simultaneous-update number                               & $K_0$                 \\
            & Maximum admissible age of cached status information              & $\overline{\Delta}_k$ \\
            & Battery capacity of sensor $k$                                   & $\overline{B}_k$      \\
            & Clipped mean of energy arrival process $ (E_{t,k})_{t=1}^\infty$ & $\lambda_k$           \\
            & Probability of requests for the physical quantity $\zeta_k$      & $\eta_k$              \\
            & Success probability of the sensor-$k$-to-edge-node channel       & $\xi_k$               \\ \bottomrule
        \end{tabular}
    \end{table}

    Consider a discrete-time energy harvesting (EH) IoT network (see Fig.~\ref{fig-system-model} and Table~\ref{tab:para-sym}) for monitoring a collection $(\zeta_k)_{k\in\mathcal{K}}$ of environmental physical quantities $\zeta_k$ (e.g., temperature or humidity), where $k\in \mathcal{K}\geqdef \integersOf{[1,K]}$.
    Each quantity $\zeta_k$ is uniquely measured by its corresponding sensor, denoted as sensor $k$.
    Each sensor $k$ is powered by an energy harvester with a rechargeable battery of size $\overline{B}_k\in \pintegers$ (units of energy).
    The sensor works in the harvest-store-use mode, that is, the energy harvested in each time slot is first stored in the battery and is available for use in the next time slot.
    By consuming one unit of energy from its battery, sensor $k$ can transmit a status update to the edge node, including the latest measurement of $\zeta_k$ and the current battery (energy) level.
    The edge node stores the latest status update in its cache.

    The entire system operates based on incoming requests from monitors.
    Suppose that a request for $\zeta_k$ is sent from a monitor to the edge node in time slot $t$.
    Based on the freshness of the cached status of $\zeta_k$, the edge node serves the request with either the full flow
    \begin{multline*}
        \text{Monitor} \xrightarrow{\text{request $\zeta_k$}} \text{Edge node} \xrightarrow{\text{command}} \text{Sensor $k$} \\
        \xrightarrow{\text{fresh status update}} \text{Edge node} \xrightarrow{\text{response}} \text{Monitor}
    \end{multline*}
    or the low-cost shortcut flow
    \[
        \text{Monitor} \xrightarrow{\text{request $\zeta_k$}} \text{Edge node} \xrightarrow{\text{response}} \text{Monitor}.
    \]
    Both flows are finished within one time slot.
    With the aid of the status cache, the edge node can always serve the request with the most recently cached status of $\zeta_k$, regardless of whether a fresh status update is received from sensor $k$.

    It is assumed that at most $K_0\le K$ sensors can send status updates to the edge node simultaneously.
    For each $k\in \mathcal{K}$, the sensor-$k$-to-edge-node link is modeled as an (independent) packet-drop channel with success probability $\xi_k$.
    All other links in the network are assumed error-free.

    \subsection{Requests and Commands}

    Let $R_{t,k}\in \mathcal{R}\geqdef \{0,1\}$ denote the request status of $\zeta_k$ in time slot $t$ (at the edge node), with $0$ and $1$ indicating the absence and presence of a request, respectively.
    All $R_{t,k}$'s are assumed to be mutually independent, and for any $k\in \mathcal{K}$, $(R_{t,k})_{t=1}^\infty$ is assumed to be independent and identically distributed (i.i.d.).
    The probability of receiving a request for $\zeta_k$ at the edge node in time slot $t$ is thus time-invariant and denoted by $\eta_k\geqdef P\{R_{t,k}=1\}$.

    Let $A_{t,k}\in \mathcal{A}\geqdef \{0,1\}$ denote whether the edge node commands sensor $k$ to send a status update in time slot $t$ (with $1$ and $0$ representing yes and no, respectively\footnote{In the sequal, we will follow this convention for all such variables with truth values.}).
    Because at most $K_0$ sensors can send status updates simultaneously, the following constraint is imposed on the commands:
    \begin{equation}
        \sum_{k=1}^K A_{t,k}
        \le K_0
        \quad \text{for all $t\ge 1$}, \label{eq:update-constraint}
    \end{equation}
    where $K_0$ is the \term{maximum simultaneous-update number}.
    The \term{maximum simultaneous-update ratio} (MSUR) is defined as $\kappa\geqdef K_0/K$.

    \subsection{Status Update}

    Upon receiving a command, sensor $k$ will send a status update to the edge node whenever its battery is not empty.
    Let $B_{t,k}$ denote the battery level of sensor $k$ at time $t$ (the beginning of time slot $t$).
    Let
    \[
        M_{t,k}
        \geqdef A_{t,k}1\{B_{t,k}\ge 1\}
    \]
    represent whether sensor $k$ sends a status update in time slot $t$.
    Since the sensor consumes one unit of energy for each status update, the battery level of sensor $k$ at time $t+1$ is
    \begin{equation}
        B_{t+1,k}
        = \upperclip{B_{t,k}-M_{t,k}+E_{t,k}}{\overline{B}_k},
    \end{equation}
    where $E_{t,k}\in \nnintegers$ denotes the amount of energy harvested by sensor $k$ in time slot $t$.
    It is assumed that all $E_{t,k}$'s are mutually independent and $(E_{t,k})_{t=1}^\infty$ is i.i.d.\ for any $k\in \mathcal{K}$.
    Let
    \begin{equation}
        \lambda_k
        \geqdef \expect \upperclip{E_{t,k}}{\overline{B}_k},
    \end{equation}
    which is the clipped (or effective) mean (see~\cite{shaviv2016universally,yang2020maximin,yang2025power}) of the energy arrival process $(E_{t,k})_{t=1}^\infty$, a quantity for characterizing the effective energy arrival rate of sensor $k$ with the battery-capacity constraint.

    Let $H_{t,k}\in \mathcal{H}\geqdef \{0,1\}$ denote whether the edge node receives the status update from sensor $k$ in time slot $t$.
    Based on the packet-drop-channel assumption, we have
    \begin{equation}
        P(\{H_{t,k}=1\}|M_{t,k}=1)
        = \xi_k.
    \end{equation}
    When $\xi_k=1$, the packet-drop channel degenerates to a noiseless channel.

    \subsection{Freshness of Response}

    The freshness of the cached status of $\zeta_k$ at the edge node at time $t$ is measured by
    \begin{equation}
        \Delta_{t,k}
        \geqdef t-U_{t,k},
    \end{equation}
    referred to as the \term{age of cached status information} (AoCSI) of $\zeta_k$, where
    \begin{equation}
        U_{t,k}
        \geqdef \max\{t': t'<t,H_{t',k}=1\} \label{eq:successful_update_time}
    \end{equation}
    denotes the time of the most recent successful status update from sensor $k$ before time $t$.
    Then, $\Delta_{t,k}$ satisfies
    \begin{equation}
        \label{eq:aocsi_evolution}
        \Delta_{t+1,k}
        =\begin{cases}
             \Delta_{t,k}+1, &\text{$H_{t,k}=0$},\\
             1, &\text{$H_{t,k}=1$}.
        \end{cases}
    \end{equation}
    The age of the status of $\zeta_k$ received at the monitor (roughly at the end of time slot $t$) is thus $\Delta_{t+1,k}$.
    The overall expected freshness of all status served to requesting monitors in time slot $t$ can be modeled as a weighted sum of the expected on-demand AoCSIs in time slot $t$, i.e.,
    \begin{equation}
        c_t(A_t)
        \geqdef \sum_{k=1}^K \alpha_k \expect\left(R_{t,k}\upperclip{\Delta_{t+1,k}}{\overline{\Delta}_k}\right) \label{eq:single_slot_cost},
    \end{equation}
    where $A_t\eqdef (A_{t,k})_{k\in\mathcal{K}}$, $\alpha_k\ge 0$ is the weight factor for $\zeta_k$, and $\overline{\Delta}_k$ is the maximum admissible AoCSI for $\zeta_k$.
    The rationale behind $\overline{\Delta}_k$ is that further increasing the value of AoCSI makes little sense when the cached status information becomes excessively stale (see, e.g., \cite{hatami2021,hatami2022}).

    \subsection{Problem Formulation}

    The central problem of status update control in this paper is how to make good decisions $A_{t,k}$ (when $R_{t,k}=1$) to optimize the freshness of the status information (in terms of~\eqref{eq:single_slot_cost}) in a time-amortized sense.
    Based on~\eqref{eq:single_slot_cost}, the total cost of the system from time slot $m$ to $n$ is
    \begin{equation}
        c_{m,n}(A_m^n)
        \geqdef \sum_{t=m}^n c_t(A_t),
    \end{equation}
    where $A_m^n\eqdef (A_t)_{t=m}^n$, and the long-term time-average cost is
    \begin{equation}
        g(A^\infty)
        \geqdef \limsup_{n\to\infty} \frac{1}{n} c_{1,n}(A^\infty),
    \end{equation}
    where $A^\infty\eqdef A_1^\infty$.
    Therefore, the problem of status update control can be roughly formulated as follows:
    \begin{problem}
        Find an optimal sequence $A^\infty$ of decisions (based on all available information) to minimize $g(A^\infty)$ subject to the constraint~\eqref{eq:update-constraint}.
    \end{problem}

    Three factors that affect the decision-making process (i.e., the sequence $A^\infty$) are the request status, the AoCSI, and the battery levels.
    However, battery levels are only available through status updates, so the edge node can only access their cached values and some relevant information.
    We call the former the \term{explicit battery state information} (explicit BSI) and the latter the \term{implicit battery state information} (implicit BSI).
    Together, these two types of information form the partial BSI (pBSI) available to the edge node.

\begin{figure}[htbp]
        \centering
        \includegraphics{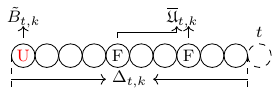}
        \caption{An example of pBSI of sensor $k$ at time $t$, where the circled U represents the successful status update, a circled F represents a failed status update, and an empty circle represents no update commanded.}
        \label{fig:bsi}
    \end{figure}

    Let
    \begin{equation}
        \tilde{B}_{t,k}
        \geqdef B_{U_{t,k},k} - 1
    \end{equation}
    denote the explicit BSI of sensor $k$ at time $t$, namely, the battery level of sensor $k$ at the time of the most recent successful status update but with one unit energy subtracted (due to the status update).
    The implicit BSI of sensor $k$ at time $t$ include two parts:
    (i) the age of explicit BSI, i.e., $\Delta_{t,k}$, which provides some information about new energy arrivals of sensor $k$ after time $U_{t,k}$;
    and (ii) the history of failed status updates (since the most recent successful status update), which provides some information about the battery levels or their changes at the time of the failed status updates.
    It is easy to see that each time slot with failed status update can be identified by the AoCSI at that time, so the history of failed status updates for sensor $k$ at time $t$ can be represented equivalently by the set
    \begin{equation}
        \overline{\mathfrak{U}}_{t,k}
        \geqdef \{\Delta_{t',k}: A_{t',k}=1, H_{t',k}=0,U_{t,k}<t'<t\}.
    \end{equation}
    The pair $(\Delta_{t,k},\overline{\mathfrak{U}}_{t,k})$ thus constitutes the implicit BSI of sensor $k$ at time $t$, and the pBSI of sensor $k$ at time $t$ is
    \begin{multline}
        O_{t,k}
        \geqdef (\tilde{B}_{t,k},\Delta_{t,k},\overline{\mathfrak{U}}_{t,k})
        \in \mathcal{O}_k
        \eqdef \bigl\{(b,\delta,\overline{U}): \\
        b\in \integersOf{[0,\overline{B}_k)}, \delta\in\pintegers, \overline{U}\subseteq\integersOf{[1,\delta)}\bigr\}. \label{eq:pbsi}
    \end{multline}
    An example of pBSI is illustrated in Fig.~\ref{fig:bsi}.
    Note that, unlike exact BSI (eBSI), the AoCSI is a sub-state of pBSI.
    Consequently, the combination of request status and pBSI provides sufficient information for optimal decision-making.

    In this paper, we focus on stationary deterministic online policies, which are time-invariant and depend only on the request status and pBSI.
    An admissible stationary (online) policy is a mapping $\pi: \prod_{k\in\mathcal{K}} \mathcal{S}_k \to \{0,1\}^K$ such that for any $(r_k,o_k)_{k\in\mathcal{K}}\in \prod_{k\in\mathcal{K}} \mathcal{S}_k$,
    \begin{equation}
        \pi_k((r_k,o_k)_{k\in\mathcal{K}})
        \le r_k \quad \text{for all $k\in\mathcal{K}$} \label{eq:request_driven_constraint}
    \end{equation}
    and
    \begin{equation}
        \sum_{k=1}^K \pi_k((r_k,o_k)_{k\in\mathcal{K}})
        \le K_0,
    \end{equation}
    where $\pi_k$ denotes the $k$th component of $\pi$, and
    \(
    \mathcal{S}_k
    \eqdef \mathcal{R}\times \mathcal{O}_k.
    \)
    The collection of all (admissible) stationary policies is denoted by $\Pi$.

    With the above definitions, the problem of status update control with pBSI can be formulated as follows:
    \begin{problem}[The pBSI Problem]
        Let
        \(
        g(\pi)
        \geqdef g(A^\infty),
        \)
        with $A_t\eqdef \pi((R_{t,k},S_{t,k})_{k\in\mathcal{K}})$.
        Let
        \(
        g^*
        \geqdef \inf_{\pi\in\Pi} g(\pi).
        \)
        Find a stationary policy $\pi$ attaining or approaching $g^*$.
    \end{problem}

    Although the above problem can be modeled as a Markov decision process (MDP) and solved using dynamic programming, its complexity is extremely high due to the infinite size of the pBSI state space $\mathcal{S}_k$.
    Furthermore, the simultaneous update constraint~\eqref{eq:update-constraint} adds another layer of complexity, as it requires coordinating updates across multiple sensors, making it challenging to find an optimal solution.
    In the next two sections, we will address the pBSI problem, starting from special cases and progressing to general cases: (i) the single-sensor case with a noiseless channel; (ii) the single-sensor case with a packet-drop channel; and (iii) the multi-sensor case.

    \section{The Noiseless Single-Sensor pBSI Problem}\label{sec:noiseless_pbsi}

    In this section, we address the pBSI problem for a single sensor with a noiseless (sensor-to-edge-node) channel, hereafter referred to as the noiseless (single-sensor) pBSI problem.
    For simplicity, we will omit the subscript $k=1$.
    Since the weight factor $\alpha$ is not relevant in the single-sensor case, we set $\alpha=1$ to simplify the analysis.
    These conventions will be applied to all single-sensor scenarios throughout this paper.

    We begin with the following crucial observation:
    The latest information about the sensor battery level is obtained from the most recent status-update attempt, regardless of its success.
    When a status update is commanded, it will always succeed if the sensor battery is not empty.
    In the only case of failure, the edge node can infer that the sensor battery level at the beginning of that time slot was zero.

\begin{figure}[htbp]
        \centering
        \includegraphics{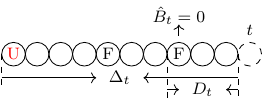}
        \caption{The relation between the inferred pBSI and the original pBSI when the most recent status update failed.}
        \label{fig:noiseless_pbsi}
    \end{figure}

    Motivated by this observation, we introduce the following reduced form of pBSI, the \term{inferred pBSI}, the triple $\tilde{O}_t$ of the \term{inferred battery level} $\hat{B}_t$, the AoCSI $\Delta_t$, and the \term{age of failure-inferred battery level} (AoFBL) $D_t$ defined by
    \begin{align}
        \tilde{O}_t
        &= (\hat{B}_t, \Delta_t, D_t) \notag \\
        &\eqdef
        \begin{cases}
        (\tilde{B}_t, \Delta_t, 0)
            , &\text{if $\overline{\mathfrak{U}}_t=\emptyset$},\\
            (0, \Delta_t, \Delta_t-\max\overline{\mathfrak{U}}_t), &\text{otherwise},
        \end{cases}\label{eq:inferred_pbsi_for_noiseless}
    \end{align}
    where $D_t=0$ indicates that $\hat{B}_t$ is not inferred from failure but obtained from the most recent status update.
    Accordingly, the \term{age of inferred battery level} (AoIBL) at time $t$ is
    \begin{equation}
        \tau(\Delta_t,D_t)
        \geqdef \Delta_t 1\{D_t=0\} + D_t, \label{eq:age_of_ibl}
    \end{equation}
    which is the elapsed time since the most recent status-update attempt.
    It can be easily shown that
    \begin{equation}
    (\hat{B}_{t+1}, D_{t+1})
        = \begin{cases}
        (\hat{B}_t, D_t+1\{D_t>0\})
              , &\text{$A_t=0$},\\
              (H_t(B_t-1), 1-H_t), &\text{$A_t=1$},
        \end{cases}\label{eq:inferred_pbsi_evolution_for_noiseless}
    \end{equation}
    where $B_t=\beta_t(\hat{B_t},\tau(\Delta_t,D_t))$, $H_t=1\{B_t>0\}$, and
    \begin{equation}
        \beta_t(\hat{b}, \hat{d})
        \eqdef \upperclip*{\hat{b} + \sum_{t=t-\hat{d}}^{t-1} E_t}{\overline{B}}. \label{eq:inferred_current_battery_level}
    \end{equation}
    Since for i.i.d.\ $(E_t)_{t=1}^\infty$, the distribution of $\beta_t(\hat{b},d)$ is independent of $t$ and hence the subscript $t$ will be omitted when only its distribution is concerned.
    Fig.~\ref{fig:noiseless_pbsi} illustrates the relation between the inferred pBSI and the original pBSI when the most recent status update failed.

    Another challenge to address is the countably infinite size of the AoCSI and AoFBL spaces, which must be adapted for the application of dynamic programming algorithms.
    Although truncating the AoCSI space to $\integersOf{[0,\overline{\Delta}]}$ seems reasonable due to the maximum admissible AoCSI in~\eqref{eq:single_slot_cost}, it introduces inaccuracies in the AoIBL (Eq.~\eqref{eq:age_of_ibl}) and consequently in the estimation of $B_t$ (Eq.~\eqref{eq:inferred_current_battery_level}).
    A similar risk also exists if we set a maximum admissible AoFBL $\overline{D}$.
    Fortunately, these potential impacts are negligible if $\lambda \gg 1/\tau_0$, where $\tau_0\eqdef \min\{\overline{\Delta}, \overline{D}\}$.
    The detailed argument is left to the reader as an easy exercise.

    Then, with the assumption~$\lambda \gg 1/\tau_0$, we can safely truncate the AoCSI and AoFBL spaces to $\integersOf{[0,\overline{\Delta}]}$ and $\integersOf{[0,\overline{D}]}$, respectively.
    The evolution laws of the truncated AoCSI and AoFBL are
    \begin{equation}
        \Delta_{t+1}
        =
        \begin{cases}
            \upperclip{\Delta_t+1}{\overline{\Delta}}, &\text{$H_{t}=0$},\\
            1, &\text{$H_{t}=1$}, \label{eq:truncated_aocsi_evolution}
        \end{cases}
    \end{equation}
    and
    \begin{equation}
        D_{t+1}
        =
        \begin{cases}
            \upperclip{D_t+1\{D_t>0\}}{\overline{D}}, &\text{$A_t=0$},\\
            1-H_t, &\text{$A_t=1$}, \label{eq:truncated_aofbl_evolution}
        \end{cases}
    \end{equation}
    respectively, where we continue using the same notations $\Delta_t$ and $D_t$ for the two truncated versions.

    We are now ready to formulate the noiseless pBSI problem as an MDP.
    It can be described by the 4-tuple $(\mathcal{S}', \mathcal{A}_s, p(s'|s,a), c(s,a))$, where
    \begin{itemize}
        \item $\mathcal{S}'$ is the state space defined by
        \begin{equation}
            \mathcal{S}'
            \eqdef \mathcal{R}\times \integersOf{[0,\overline{B})}\times \integersOf{[1,\overline{\Delta}]}\times \integersOf{[0,\overline{D}]},
        \end{equation}
        and a state $s\in \mathcal{S}'$ is a $4$-tuple $(r,\hat{b},\delta,d)$ of the request $r$, inferred battery level $\hat{b}$, AoCSI $\delta$, and AoFBL $d$;
        \item $\mathcal{A}_s\eqdef \integersOf{[0,r]}$ is the set of allowable actions in state $s=(r,\hat{b},\delta,d)$, and the whole action space is $\mathcal{A}=\{0,1\}$;
        \item $p(s'|s,a)$ is the transition probability from the current state $s=(r,\hat{b},\delta,d)$ to the next state $s'\eqdef (r',\hat{b}',\delta',d')$ given the action $a$, defined by
        \begin{subequations}
            \begin{multline}
                p(s'|s,0)
                = \rho(r') 1\{\hat{b}'=\hat{b}\} 1\left\{\delta'=\upperclip{\delta+1}{\overline{\Delta}}\right\} \\
                \cdot 1\left\{d'=\upperclip{d+1\{d>0\}}{\overline{D}}\right\},
            \end{multline}
            \begin{multline}
                p(s'|s,1)
                = \rho(r') P\{\beta(\hat{b},\tau(\delta,d))=\hat{b}'+1\} \\
                \cdot 1\{\delta'=1,d'=0\}
                + \epsilon \rho(r') 1\{\hat{b}'=0\} \\
                \cdot 1\left\{\delta'=\upperclip{\delta+1}{\overline{\Delta}},d'=1\right\},
            \end{multline}
        \end{subequations}
        with
        \begin{equation}
            \rho(r)
            \eqdef \eta 1\{r=1\} + (1-\eta) 1\{r=0\}
        \end{equation}
        and
        \begin{equation}
            \epsilon
            \eqdef P\{\beta(\hat{b},\tau(\delta,d))=0\}
        \end{equation}
        being the probability of the request status and the probability that the battery is empty at the time of action, respectively;
        \item $c(s,a)$ is the cost of taking action $a$ in state $s=(r,\delta,\hat{b},d)$ (see~\eqref{eq:single_slot_cost}), and is calculated by
        \begin{subequations}
            \begin{equation}
                c(s,0)
                = r\upperclip{\delta+1}{\overline{\Delta}},
            \end{equation}
            \begin{equation}
                c(s,1)
                = \epsilon \upperclip{\delta+1}{\overline{\Delta}} + 1-\epsilon.
            \end{equation}
        \end{subequations}
    \end{itemize}

    Using the above formulation, the optimal policy for the noiseless pBSI problem can be obtained through the relative value iteration (RVI) algorithm (see, e.g., \cite[Sec.~8.5.5]{puterman}).
    The RVI algorithm is presented in Algorithm~\ref{PI}.
    \begin{algorithm}
        \caption{Relative Value Iteration (RVI) Algorithm That Obtains the Optimal Policy $\pi^*$ for the Noiseless pBSI Problem}
        \label{PI}
        \begin{algorithmic}[1]
            \State Initialization: $V(s)\gets 0, h(s)\gets 0,\;\forall s\in\mathcal{S}'$, choose a reference state $s_{\text{ref}}$ and a small $\vartheta>0$.
            \Repeat
                \State $V'(s)\gets V(s),\;\forall s\in\mathcal{S}'$
                \For{$s\in \mathcal{S}'$}
                    \State $V(s)\gets \min_{a\in\mathcal{A}_s} ( c(s,a) + \sum_{s'\in\mathcal{S}'} p(s'|s,a) h(s') )$
                \EndFor
                \State $\theta\gets \max_{s\in\mathcal{S}'} (V(s)-V'(s)) - \min_{s\in\mathcal{S}'} (V(s)-V'(s))$
                \State $h(s)\gets V(s)-V(s_{\text{ref}}),\;s\in\mathcal{S}'$
            \Until{$\theta<\vartheta$}
            \State $\pi^*(s)\gets \arg\min_{a\in\mathcal{A}_s} ( c(s,a)$
            \State $\hspace{9em} + \sum_{s'\in\mathcal{S}'} p(s'|s,a) h(s') ),\;\forall s\in\mathcal{S}'$
        \end{algorithmic}
    \end{algorithm}
    A key practical concern is the time complexity of the algorithm, which depends on the size of the state space $\mathcal{S}'$.
    Therefore, we proceed to analyze the size of $\mathcal{S}'$.

    \begin{proposition}
        \label{pr:noiseless_pbsi_state_space_size}
        The effective size of the state space  $\mathcal{S}'$ is
        $2\overline{B}\overline{\Delta} + 2\overline{D}(\overline{\Delta}-\tau_1+1) + (\tau_1 - 1)(\tau_1 - 2)$, where $\tau_1\eqdef \min\{\overline{\Delta},\overline{D}+1\}$.
    \end{proposition}

    \begin{IEEEproof}
        At first, we have
        \begin{equation}
            |\mathcal{S}'| = 2 |\mathcal{S}'_0|, \label{eq:noiseless_pbsi_state_space_size.1}
        \end{equation}
        where $\mathcal{S}'_0\eqdef \integersOf{[0,\overline{B})}\times \integersOf{[1,\overline{\Delta}]}\times \integersOf{[0,\overline{D}]}$.

        Next, we estimate the effective size of the sub-state space $\mathcal{S}'_0$.
        Observe that, for any $(\hat{b},\delta,d)\in \mathcal{S}'_0$, we must have $\hat{b}=0$ if $d>0$.
        Therefore, the effective size of $\mathcal{S}'_0$ is at most
        \begin{equation}
            \left|\integersOf{[0,\overline{B})}\times \integersOf{[1,\overline{\Delta}]} \times \{0\}\right| + \left|\{0\}\times \integersOf{[1,\overline{\Delta}]}\times \integersOf{[1,\overline{D}]}\right|
            = \overline{\Delta} (\overline{B} + \overline{D}). \label{eq:noiseless_pbsi_state_space_size.2}
        \end{equation}

        This upper bound can be made exact by eliminating the following infeasible states.
        Note that any state $(0,\delta,d)$ with $\delta<\overline{\Delta}$ is infeasible if $\delta \le d$.
        The number of all such infeasible states is
        \begin{align}
            &\left|\{(\delta, d)\in \integersOf{[1,\overline{\Delta})}\times \integersOf{[1,\overline{D}]}: \delta \le d\}\right| \notag \\
            &\qquad = \sum_{\delta=1}^{\overline{\Delta}-1} \lowerclip{\overline{D}-\delta+1}{0}
            = \sum_{\delta=1}^{\tau_1-1} (\overline{D}-\delta+1) \notag \\
            &\qquad = \frac{(\tau_1-1)(2\overline{D}+2-\tau_1)}{2}. \label{eq:noiseless_pbsi_state_space_size.3}
        \end{align}
        Combining~\eqref{eq:noiseless_pbsi_state_space_size.3} with~\eqref{eq:noiseless_pbsi_state_space_size.1} and~\eqref{eq:noiseless_pbsi_state_space_size.2} thus gives the effective size of $\mathcal{S}'$:
        \begin{multline*}
            2\overline{\Delta} (\overline{B} + \overline{D}) - (\tau_1-1)(2\overline{D}+2-\tau_1) \\
            = 2\overline{B}\overline{\Delta} + 2\overline{D}(\overline{\Delta}-\tau_1+1) + (\tau_1 - 1)(\tau_1 - 2).
        \end{multline*}
    \end{IEEEproof}

    Proposition~\ref{pr:noiseless_pbsi_state_space_size} shows that, although an additional variable, AoFBL, is introduced to track the pBSI, the effective size of the state space $\mathcal{S}'$ is only slightly larger than that of the eBSI problem, which is $2\overline{B}\overline{\Delta}$ (see~\cite{hatami2022}).
    As an easy consequence of Proposition~\ref{pr:noiseless_pbsi_state_space_size}, we have the following corollary on the time complexity of Algorithm~\ref{PI}.

    \begin{corollary}
        \label{co:noiseless_pbsi_vi_complexity}
        If the sparsity of transition probabilities is leveraged, the per-iteration time complexity of Algorithm~\ref{PI} is $\mathrm{O}(\overline{B}\overline{\Delta} (\overline{B} + \overline{D}))$.
    \end{corollary}

    \begin{IEEEproof}
        It is evident that the computational complexity of each iteration in Algorithm~\ref{PI} is $\mathrm{O}(T|\mathcal{S}'||\mathcal{A}|)$ when measured in terms of the number of multiplication operations required, where
        \[
            T
            \eqdef \max_{s\in\mathcal{S}',a\in\mathcal{A}} \left|\{s':p(s'|s,a)>0\}\right|
            = \mathrm{O}(\overline{B}|\mathcal{R}|).
        \]
        From Proposition~\ref{pr:noiseless_pbsi_state_space_size}, it follows that the effective size of the state space $\mathcal{S}'$ is $\mathrm{O}(\overline{\Delta} (\overline{B}+\overline{D}))$.
        Therefore, the per-iteration time complexity of Algorithm~\ref{PI} is $\mathrm{O}(\overline{B}\overline{\Delta} (\overline{B}+\overline{D}))$.
    \end{IEEEproof}

    By a similar argument, it can be shown that the (per-iteration) time complexity of RVI in the eBSI case with general energy arrival distributions is $\mathrm{O}(\overline{B}^2\overline{\Delta})$.
    Thus, the time complexity of Algorithm~\ref{PI} is comparable to that of the eBSI case if $\overline{D}/\overline{B}$ is not too large.
    For comparison, the algorithm proposed in~\cite[Alg.~2]{hatami2024status} for solving the noiseless pBSI problem under Bernoulli energy arrival processes has a time complexity of $\mathrm{O}(M\overline{B}^2\overline{\Delta})$, where $M$ is a belief-space truncation parameter analogous to $\overline{D}$.
    For comparable values of $\overline{D}$ and $M$, the time complexity of Algorithm~\ref{PI} is significantly lower than that of~\cite[Alg.~2]{hatami2024status}.

    \section{The General Single-Sensor pBSI Problem}\label{sec:general_pbsi}

    In this section, we address the single-sensor pBSI problem for a general packet-drop channel, referred to as the general (single-sensor) pBSI problem. Although the approach in Section~\ref{sec:noiseless_pbsi} cannot be directly extended to the case of general packet-drop channels, valuable insights can be drawn from it. The solution to the noiseless pBSI problem in Section~\ref{sec:noiseless_pbsi} effectively reduces the size of the pBSI state space by introducing the concept of inferred pBSI, which simplifies the state representation and improves the efficiency of the dynamic programming algorithm.
    However, this approach is not fully effective, as it relies on a general value iteration algorithm, whose time complexity remains high due to the large state space (Corollary~\ref{co:noiseless_pbsi_vi_complexity}). These insights motivate the search for a similar inferred pBSI and the identification of a small set of crucial states for the general pBSI problem.
    Based on the inferred pBSI and an approximation of the crucial-state value function, we will design a low-complexity near-optimal status update control scheme.

    Before presenting the approach to the general pBSI problem, we first establish a lower bound on the performance of all policies, including those with eBSI, pBSI, or nBSI.

    \begin{theorem}
        \label{th:lower_bound}
        For $\xi \ge \underline{\xi} \geqdef 1/(\overline{\Delta}-1/2)$, the expected time-average on-demand AoCSI under any policy (with eBSI, pBSI, or nBSI) is bounded below by
        \begin{multline}
            \Theta(\lambda,\eta,\xi) \\
            \quad \geqdef
            \begin{dcases}
                \frac{\eta}{2} + \frac{\eta}{2\xi\upperclip{\lambda}{\eta}} \\
                \quad{} - \frac{1-\eta}{\xi} \left(1 - \frac{\upperclip{\lambda}{\eta}}{2\eta}\right), &\text{$\lambda \ge \lambda_0(\eta, \xi)$}, \\
                \eta \left[\overline{\Delta} - \frac{\lambda\xi}{2}\left(\overline{\Delta}-\frac{1}{2}\right)^2\right] \\
                \quad{} - \lambda (1-\eta) \left(\overline{\Delta} - \frac{1}{2\xi} - \frac{1}{2}\right), &\text{$\lambda < \lambda_0(\eta, \xi)$},
            \end{dcases} \label{eq:lower-bound}
        \end{multline}
        where
        \begin{equation}
            \lambda_0(\eta, \xi)
            \geqdef \frac{1}{\left(\overline{\Delta}-\dfrac{1}{2}\right)\xi + \dfrac{1}{\eta} - 1}. \label{eq:lambda_0}
        \end{equation}
        \proofref{in Section~\ref{proof:lower_bound}}
    \end{theorem}

    \begin{remark}
        Inspecting the proof of Theorem~\ref{th:lower_bound} reveals that the lower bound $\Theta(\lambda,\eta,\xi)$ is relatively tight when $\eta=1$ or $\xi=1$.
        However, as both $\eta$ and $\xi$ decrease, the lower bound $\Theta(\lambda,\eta,\xi)$ becomes looser.
        It is thus of interest to find a tighter lower bound for the general case.
    \end{remark}

    \subsection{The Generalized Inferred pBSI}

    The core idea behind the inferred pBSI~\eqref{eq:inferred_pbsi_for_noiseless} in the noiseless case is to find a simplified representation of the implicit BSI that retains information about new energy arrivals not consumed by previous failed status updates (if any).
    However, this approach cannot be directly applied to general packet-drop channels, as the success of a status update is no longer guaranteed when the battery is not empty, and the edge node cannot accurately infer the battery level from failed updates.
    Nevertheless, we make the following key observations regarding the pBSI $O_t=(\tilde{B}_t,\Delta_t,\overline{\mathfrak{U}}_t)$ (Eq.~\eqref{eq:pbsi}):

    (1) When the number $|\overline{\mathfrak{U}}_t|$ of update failures is less than or equal to the explicit BSI $\tilde{B}_t$, the current battery level is $\tilde{B}_t-|\overline{\mathfrak{U}}_t|$ plus the harvested energy since the most recent successful status update, clipped by the battery capacity $\overline{B}$.
    In this case, the history of update failures does not provide additional information about the battery level, and hence the pBSI can be equivalently represented by $(\tilde{B}_t-|\overline{\mathfrak{U}}_t|,\Delta_t,\emptyset)$.
    This simplification reduces the complexity of the state representation by eliminating the need to track the history of update failures.

    (2) On the other hand, once the number of update failures exceeds the explicit BSI, the evolution of the distribution of the battery level becomes more complex.
    Each of the last $|\overline{\mathfrak{U}}_t-\tilde{B}_t|$ failed status updates may or may not have reduced the battery level, and this cannot be determined solely from the status update failure because in packet-drop channels, a failed update can result from either an empty battery or packet loss.
    To avoid the need to track the history of update failures, we instead focus on the expected battery level given the pBSI.

    \begin{proposition}
        \label{pr:expected_battery_level}
        Suppose that $\overline{B}=\infty$ and $O_t=(\tilde{B}_t,\Delta_t,\overline{\mathfrak{U}}_t) = o\eqdef (0,\delta,\overline{U})$.

        (1) If $A_t=0$ (no update), then $O_{t+1} = (0,\delta+1,\overline{U})$ and
        \begin{equation}
            \expect(B_{t+1}|O_{t+1}=(0,\delta+1,\overline{U}))
            = \expect(B_t|O_t=o,A_t=0) + \lambda. \label{eq:expected_battery_level_for_no_update}
        \end{equation}

        (2) If $A_t=1$ but $H_t=0$ (update failure), then $O_{t+1} = (0,\delta+1,\overline{U}\cup\{\delta\})$ and
        \begin{multline}
            \expect(B_{t+1}|O_{t+1}=(0,\delta+1,\overline{U}\cup\{\delta\})) \\
            = \frac{(1-\xi)(\expect(B_t|O_t=o,A_t=1)-\theta_t)}{1-\theta_t\xi} + \lambda, \label{eq:expected_battery_level_for_update_failure}
        \end{multline}
        where $\theta_t\geqdef P\{B_t>0|O_t=o,A_t=1\}$.

        (3) If $A_t=1$ and $H_t=1$ (update success), then $O_{t+1} = (B_t-1,1,\emptyset)$ and
        \begin{equation}
            \expect(B_{t+1}|O_{t+1}=(B_t-1,1,\emptyset))
            = B_t - 1 + \lambda. \label{eq:expected_battery_level_for_update_success}
        \end{equation}
    \end{proposition}

    \begin{IEEEproof}[Sketch of Proof]
        The cases (1) and (3) are trivial.
        The case (2) can be proved with the following observations:
        \begin{gather*}
            \begin{aligned}
                &\expect(B_{t+1}|O_t=o,B_t>0,A_t=1,H_t=0) \\
                &\quad = \expect(B_{t+1}|O_t=o,B_t>0,A_t=1) \\
                &\quad = \expect(B_t-1+E_t|O_t=o,B_t>0,A_t=1) \\
                &\quad = \frac{\expect(B_t|O_t=o,A_t=1)}{\theta_t} - 1 + \lambda,
            \end{aligned} \\
            \begin{multlined}
                \expect(B_{t+1}|O_t=o,B_t=0,A_t=1,H_t=0) \\
                = \expect(E_t|O_t=o,B_t=0,A_t=1)
                = \lambda,
            \end{multlined} \\
            P\{B_t>0,H_t=0|O_t=o,A_t=1\}
            = \theta_t(1-\xi), \\
            P\{B_t=0,H_t=0|O_t=o,A_t=1\}
            = 1-\theta_t, \\
            P\{H_t=0|O_t=o,A_t=1\}
            = 1-\theta_t\xi.
        \end{gather*}
    \end{IEEEproof}

    \begin{remark}
        \label{re:generalized_aofbl}
        Due to the possible dependence of $A_t$ on $B_t$, it is not easy to compute $\expect(B_t|O_t=o,A_t=0)$, $\expect(B_t|O_t=o,A_t=1)$, and $\theta_t$ accurately.
        In practice, we just use the following approximations:
        \begin{align}
            \expect(B_t|O_t=o,A_t=0)
            &\approx \expect(B_t|O_t=o) \\
            \expect(B_t|O_t=o,A_t=1)
            &\approx \expect(B_t|O_t=o) \\
            \theta_t
            &\approx P\{B_t>0|O_t=o\} \notag\\
            &\approx 1 - (1-p_1)^{\expect(B_t|O_t=o)/\lambda}, \label{eq:approximated_theta}
        \end{align}
        where $p_1\eqdef P\{E_1>0\}$.
        To make these approximations precise, we need to reduce the dependence of $A_t$ on $B_t$.
        Therefore, we will minimize the use of battery level information in decisions when $\tilde{B}_t=0$.
        By considering the equivalent time for energy harvesting, $\expect(B_t|O_t)$ can be expressed as
        \(
        D_t
        \eqdef \expect(B_t|O_t)/\lambda,
        \)
        a generalization of the AoFBL (defined in~\eqref{eq:inferred_pbsi_for_noiseless}).
    \end{remark}

    Based on the above discussions, we can define the (generalized) inferred pBSI as the triple $\tilde{O}_t$ of the inferred battery level $\hat{B}_t$, the AoCSI $\Delta_t$, and the (generalized) AoFBL $D_t$, namely,
    \begin{align}
        \tilde{O}_t
        &= (\hat{B}_t, \Delta_t, D_t) \notag \\
        &\eqdef
        \begin{cases}
        (\tilde{B}_t-|\overline{\mathfrak{U}}_t|, \Delta_t, 0)
            , &\text{if $|\overline{\mathfrak{U}}_t|\le \tilde{B}_t$},\\
            \left(0, \Delta_t, \dfrac{\expect(B_t|O_t)}{\lambda}\right), &\text{otherwise}.
        \end{cases}\label{eq:inferred_pbsi_for_deletion}
    \end{align}
    The evolution of $\Delta_t$ is given by~\eqref{eq:aocsi_evolution} and the evolution of $\hat{B}_t$ and $D_t$ is given by
    \begin{align}
        \hat{B}_{t+1}
        =
        \begin{cases}
            \hat{B}_t, &\text{$A_t=0$},\\
            H_t(B_t-1)+(1-H_t)\lowerclip{\hat{B}_t-1}{0}, &\text{$A_t=1$},
        \end{cases}\label{eq:inferred_battery_level_evolution}
    \end{align}
    and
    \begin{align}
        D_{t+1}
        =
        \begin{cases}
            D_t+1\{D_t>0\}, &\text{$A_t=0$},\\
            1\{\hat{B}_t=0\}(1-H_t) \\
            \quad \cdot \left[\dfrac{(1-\xi)(\tau(\Delta_t,D_t)-\theta_t/\lambda)}{1-\theta_t\xi} + 1\right], &\text{$A_t=1$},
        \end{cases}\label{eq:aofbl_evolution}
    \end{align}
    respectively, where $\tau(\Delta_t,D_t)$ is the AoIBL defined by~\eqref{eq:age_of_ibl} and $\theta_t$ is given by~\eqref{eq:approximated_theta}.
    Note that in the noiseless case ($\xi=1$), Eqs.~\eqref{eq:inferred_battery_level_evolution} and~\eqref{eq:aofbl_evolution} just degenerate to~\eqref{eq:inferred_pbsi_evolution_for_noiseless}.
    Unlike the noiseless case, the AoFBL $D_t$ is in general a real number in $\{0\}\cup [1,+\infty)$.
    Moreover, thanks to the low-complexity scheme to be introduced, both the AoCSI and the AoFBL, if only used for decision-making, no longer need a truncation like~\eqref{eq:truncated_aocsi_evolution} and~\eqref{eq:truncated_aofbl_evolution}.

    \subsection{The Current-Next Updating Policy}

    In this subsection, we will choose a subset of states as crucial states and then derive a closed-form updating rule based on the assumed crucial-state value function.

    A good candidate for the crucial states is the set of states where the AoCSI is one.
    In this case, the explicit BSI is the freshest among all possible states, because a successful status update has just been finished.
    This motivates the concept of post-update states.

    An inferred pBSI state with the form $\tilde{o}=(\hat{b},1,0)$ is called a \term{post-update pBSI state}, where $\hat{b}\in\integersOf{[0,\overline{B})}$.
    Correspondingly, any state $s=(r,\hat{b},1,0)$ is called a \term{post-update state}, where $r$ is the request status.
    According to~\cite[Eq.~(8.2.5)]{puterman}, the $N$-step total value of a post update state (under an optimal policy) is
    \begin{equation}
        Ng^*+h(r,\hat{b},1,0)+\mathrm{o}(1),
    \end{equation}
    where $g^*$ is the optimal cost and $h(r,\hat{b},1,0)$ is called the bias function or the relative-value function.
    Since the request status is i.i.d. and is independent of the history of all past states and actions, we further consider the expectation of the bias function with respect to the request status:
    \begin{equation}
        \tilde{h}(\hat{b})
        \geqdef \expect(h(R_1,\hat{b},1,0)),
    \end{equation}
    referred to as the \term{post-update value function}.
    Therefore, the expected $N$-step total value of a post-update pBSI state (with a random request status) is
    \begin{equation}
        Ng^*+\tilde{h}(\hat{b})+\mathrm{o}(1).
    \end{equation}

    Next, we derive a closed-form updating rule based on the post-update value function.
    Recalling that the system is allowed to update only when there is a request, we suppose that the current state $S_t=s=(1,\hat{b},\delta,d)$.
    There are only two possible actions: update ($A_t=1$) or do not update ($A_t=0$).
    It suffices to compute the $N$-step total values for both actions and select the action with the smaller value.

    For better understanding the main trick to be used, we begin with the simplest case where both the request probability $\eta$ and the success probability $\xi$ are equal to one.
    It is also assumed that $\hat{b}>0$ and $\delta<\overline{\Delta}$.
    This implies that the sensor's battery is not empty and the cached status information is not so stale that skipping an update may still be worth considering.

    In this case, if $A_t=1$, then the system reaches a post-update state $s_{t+1}=(1,B_t-1,1,0)$ at time $t+1$.
    The $N$-step total value for $A_t=1$ (and the subsequent optimal actions) is
    \begin{equation}
        V_N(s|A_t=1)
        \approx 1 + (N-1)g^* + \expect\tilde{h}(B_t-1). \label{eq:vn_1}
    \end{equation}
    On the other hand, if $A_t=0$, then it is unknown when the system reaches a post-update state, which is determined by the subsequent actions, more specifically, the time of the next update.
    Let $T'\ge t+1$ be the time of the next update, which may be random.
    Then, the $N$-step total value for $A_t=0$ is
    \begin{equation}
        V_N(s|A_t=0)
        = \sum_{i=1}^\infty P\{T'=t+i\} V_N(s|A_t^{t+i}=0^i1),
    \end{equation}
    where $A_t^{t+i}\eqdef (A_j)_{j=t}^{t+i}$, and $0^i1$ denotes a sequence of $i$ zeros followed by a one.
    It is challenging to estimate $V_N(s|A_t^{t+i}=0^i1)$ for all $i$.
    Instead, we only consider the $N$-step total value for $A_t^{t+1}=01$ (and the subsequent optimal actions).
    We have
    \begin{equation}
        V_N(s|A_t^{t+1}=01)
        \approx \delta + 2 + (N-2)g^* + \expect\tilde{h}(B_t+E_t-1). \label{eq:vn_01}
    \end{equation}
    Let
    \begin{align}
        \Delta V(s)
        &\eqdef \lim_{N\to\infty} (V_N(s|A_t=1) - V_N(s|A_t^{t+1}=01)) \notag \\
        &\approx \expect(\tilde{h}(B_t-1) - \tilde{h}(B_t+E_t-1)) + g^* - (\delta + 1) \notag \\
        &\relvar{(a)}{\approx} \tilde{h}(\hat{b}+\lambda\delta-1) - \tilde{h}(\hat{b}+\lambda(\delta+1)-1) + g^* \notag \\
        &\quad - (\delta + 1), \label{eq:delta_v_basic}
    \end{align}
    where (a) follows from the empirical observation that under smooth interpolation, $\tilde{h}''(x)\ll |\tilde{h}'(x)|$.\footnote{%
        While interchanging the order of the expectation and the function $\tilde{h}(x)$ introduces errors, these errors approximately cancel each other out in the approximation of $\Delta V(s)$.
        By performing a second-order Taylor expansion and leveraging the independence between $B_t$ and $E_t$, it can be shown that interchanging the order of the expectation and the function $\tilde{h}(x)$ has a negligible effect on the approximation of $\Delta V(s)$ if $\tilde{h}''(x)\ll |\tilde{h}'(x)|$ and $\var(E_t)$ is not large.
        In the subsequent analysis, we will directly apply this approximation to estimate the N-step total value for a given strategy.%
    }
    Note that this function is independent of $t$.

    We suppose that $\Delta V(s)$ (or its approximation) has a threshold-based structure, that is, if $\Delta V(s)<0$, then $\Delta V(s')<0$ for any $s'=(1,\hat{b}',\delta',0)$ with $\hat{b}'\ge \hat{b}$ and $\delta'\ge \delta$.
    If $\Delta V(s)>0$, then it is clear that updating at time $t$ is not optimal.
    If $\Delta V(s)<0$, then $\Delta V(1,\hat{b},\delta+i,d)<0$ for all $i\ge 0$ (threshold-based structure).
    Equivalently, we have $V_N(s|A_t^{t+i}=0^i1) < V_N(s|A_t^{t+i+1}=0^{i+1}1)$ for all $i\ge 0$.
    Therefore,
    \begin{align*}
        V_N(s|A_t=1)
        &< V_N(s|A_t^{t+1}=01) \\
        &< \sum_{i=1}^\infty P\{T'=i\} V_N(s|A_t^{t+i}=0^i1) \\
        &= V_N(s|A_t=0),
    \end{align*}
    which implies that the optimal action at time $t$ is to update.

    In summary, if the derived $\Delta V(s)$ has a threshold-based structure, then the optimal action at time $t$ is to update if $\Delta V(s)<0$ and not to update if $\Delta V(s)>0$.
    As for the equal-value case ($\Delta V(s)=0$), both actions may be chosen, and we choose not to update for simplicity.
    In other words, in order to determine whether to update, it suffices to compare the $N$-step total values under two strategies: (i) update in the current time slot (with request), denoted as $\Gamma_1$; and (ii) wait until the next time slot with request, denoted as $\Gamma_2$.
    Note that these two strategies are essentially non-stationary policies, which are assumed to switch to the optimal stationary policy after the first successful update.

    The situation for $\xi<1$ becomes more complex because the update may fail even if the sensor battery is not empty.
    We thus modify the above updating rule by replacing the single updates in $\Gamma_1$ and $\Gamma_2$ with consecutive updates at time slots with request until a successful update occurs.

    The above discussion does not cover the case $\hat{b}=0$.
    In this case, we are not sure if the current sensor battery is not empty, and the battery level information from $d$ (or $\delta$ if $d=0$) may be highly unreliable.
    Due to Remark~\ref{re:generalized_aofbl}, we also need to minimize the use of battery level information in decisions when $\hat{b}=0$.
    We thus take a simple ``extension'' strategy.
    Given a state $s=(1,0,\delta,d)$, if the $\Delta V(s)$-based decision for $s'=(1,1,\delta,0)$ is not to update, then we choose not to update (for the state $s$); otherwise, we choose to update.

    Then, we have the following formal definition of the updating policy.

    \begin{definition}
        \label{df:cn_policy}
        For the state $s=(r,\hat{b},\delta,d)$, the \term{current-next (CN) updating policy} is defined as
        \begin{equation}
            \cnu(s)
            \geqdef r 1\{\Delta V(1,\lowerclip{\hat{b}}{1},\delta,0) < 0\},
        \end{equation}
        where
        \begin{equation}
            \Delta V(s)
            \eqdef \lim_{N\to\infty} (V_N(s|\Gamma_1) - V_N(s|\Gamma_2)),
        \end{equation}
        and $V_N(s|\Gamma_i)$ denotes the $N$-step total value for the strategy $\Gamma_i$.
    \end{definition}

    The next proposition gives the estimations of $V_N(s|\Gamma_1)$ and $V_N(s|\Gamma_2)$.

    \begin{theorem}
        \label{th:n-value}
        For any state $s=(r,\hat{b},\delta,0)$ with $\hat{b}\ge 1$,
        \begin{align}
            &V_N(s|\Gamma_1)
            \approx \sum_{i=1}^{\bar{n}} \xi(1-\xi)^{i-1} \hat{V}_N(b_0,\delta,i) \notag \\
            &\quad + (1-\xi)^{\bar{n}} \bigl[ \psi_1(\delta, \bar{n}) + \tilde{V}_{N-\phi_0(\bar{n})}(b_1, \psi_0(\delta,\bar{n}), \phi_1(b_1)) \bigr], \label{eq:n-value.1} \\
            &V_N(s|\Gamma_2)
            \approx \sum_{i=1}^{\bar{n}} \xi(1-\xi)^{i-1} \hat{V}_N\left(b_0'-\frac{\lambda}{\eta}+1,\delta,i+1\right) \notag \\
            &\quad + (1-\xi)^{\bar{n}} \bigl[ \psi_1(\delta, \bar{n}+1) \notag \\
            &\quad + \tilde{V}_{N-\phi_0(\bar{n}+1)}(b_2, \psi_0(\delta,\bar{n}+1), \phi_1(b_2)) \bigr], \label{eq:n-value.2}
        \end{align}
        where $b_0\eqdef \upperclip{\hat{b}+\lambda \delta}{\overline{B}}$, $\bar{n}\eqdef \lfloor b_0 \rfloor$,
        \begin{gather}
            b_0'
            =\upperclip*{b_0+\frac{\lambda}{\eta}}{\overline{B}}, \label{eq:b_0_adjustment}\\
            b_1
            \eqdef b_0-\bar{n}+\lambda\phi_0(\bar{n}),
            \quad
            b_2
            \eqdef b_0'-\bar{n}+\lambda\phi_0(\bar{n}), \label{eq:b_2}\\
            \psi_0(\delta, i)
            \eqdef \delta + \phi_0(i), \\
            \psi_1(\delta, i)
            \eqdef \sum_{j=1}^i \upperclip{\psi_0(\delta, j)}{\overline{\Delta}} , \label{eq:psi_1} \\
            \phi_0(i)
            \eqdef 1 + \frac{i-1}{\eta}, \label{eq:phi_0}\\
            \phi_1(b)
            \eqdef \max\left\{\frac{1}{\xi\lambda} - \frac{b}{\lambda}, \frac{1}{\xi\eta}-\frac{1}{\eta}+1\right\},
        \end{gather}
        and $\hat{V}(b,\delta,i)$ and $\tilde{V}(b,\delta,x)$ are given in Propositions~\ref{pr:kth_request_estimation} and~\ref{pr:xth_step_estimation}, respectively.
        \proofref{in Section~\ref{proof:n-value}}
    \end{theorem}

    \begin{proposition}
    [The $i$-th-Request-Success Estimation]
        \label{pr:kth_request_estimation}
        If $R_t=1$, $\expect B_t = b$, and $\Delta_t=\delta$, then for any $i\le b$, the $N$-step total value under strategy $\Gamma_1$ given that the first successful update occurs at the $i$-th time slot with request can be estimated by
        \begin{multline}
            \hat{V}_N(b,\delta,i)
            \approx 1 + \psi_1(\delta,i-1) \\
            + (N-\phi_0(i))g^* + \tilde{h}(b-i+\lambda\phi_0(i)-\lambda),
        \end{multline}
        where $\psi_1(\delta,i)$ and $\phi_0(i)$ are defined by~\eqref{eq:psi_1} and~\eqref{eq:phi_0}, respectively.
        \proofref{in Section~\ref{proof:n-value}}
    \end{proposition}

    \begin{proposition}
    [The $x$-th-Step-Success Estimation]
        \label{pr:xth_step_estimation}
        If $\expect B_t = b$ and $\Delta_t=\delta$, then for any $x\ge 1$, the $N$-step total value under strategy $\Gamma_1$ given that the first successful update occurs at the $x$-th step can be estimated by
        \begin{multline}
            \tilde{V}_N(b,\delta,x)
            \approx 1 + \eta\psi_2(\delta,x-1) \\
            + (N-x)g^* + \tilde{h}(\lowerclip{b-\phi_2(x)+\lambda x-\lambda}{0}),
        \end{multline}
        where
        \begin{align}
            \psi_2(\delta,x)
            &\eqdef \psi_1(\delta,\lfloor x \rfloor) + (x-\lfloor x\rfloor) \upperclip{\psi_0(\delta,\lceil x\rceil)}{\overline{\Delta}}, \label{eq:psi_2}\\
            \phi_2(x)
            &\eqdef 1 + \eta(x - 1). \label{eq:phi_2}
        \end{align}
        \proofref{in Section~\ref{proof:n-value}}
    \end{proposition}

    In order to use the CN policy, we need to estimate the post-update value function $\tilde{h}(\hat{b})$, which is the task of the next subsection.

    \subsection{Estimation of the Post-Update Value Function}\label{subsec-srvi}

    In order to design a low-complexity algorithm to estimate the post-update value function $\tilde{h}(\hat{b})$, we take an online-offline hybrid approach.
    Roughly speaking, we convert an $mn$-time-slot MDP into a simplified $m$-block MDP, with each block corresponding to $n$ time slots.
    To simulate the original MDP under the optimal policy, we have the following ideas about the states, actions, and rewards of the new MDP:
    \begin{enumerate}
        \item We assume that the $i$-th block starts with a post-update pBSI state $\tilde{O}_i=(\hat{B}_i,1,0)$, where $\hat{B}_i\le \overline{B}-1$ is the inferred battery level.
        In addition, we assume non-causal knowledge $\check{E}_i$ of the total energy arrival during the period from the last time slot of the previous block to the second-to-last time slot of the current block.
        This is referred to as \term{block energy arrival}.

        \item To further reduce the complexity of the new MDP, we assume that the block energy arrival $\check{E}_i$ follows a three-point distribution based on the mean $\lambda$ and standard deviation $\sigma$ of the battery-capacity-clipped energy arrival process:
        \begin{multline}
            P_{\check{E}_i}(\check{e})
            \geqdef \hat{p}_{-1} 1\{\check{e}=\hat{e}_{-1}\} \\
            + \left(\frac{17}{18}-\hat{p}_{-1}\right) 1\{\check{e}=\hat{e}_0\} + \frac{1\{\check{e}=\hat{e}_{1}\}}{18}, \label{eq:three_point_distribution}
        \end{multline}
        where
        \begin{subequations}
            \begin{gather*}
                \hat{e}_0
                \geqdef n\lambda,\;
                \hat{e}_{-1}
                \geqdef \lowerclip{\hat{e}_0 - 3\sigma\sqrt{n}}{1},\;
                \hat{e}_{1}
                \geqdef \hat{e}_0 + 3\sigma\sqrt{n}, \\
                \hat{p}_{-1}
                \geqdef \frac{\sigma\sqrt{n}}{6(n\lambda-\hat{e}_{-1})}.
            \end{gather*}
        \end{subequations}
        Note that the left point $\check{e}_{-1}$ is adjusted to ensure that there is at least one unit of energy arrival in each block.

        \item We assume that the effective block action is the block update number, say $N_{\text{u}}$, which cannot be larger than the block request number, say $N_{\text{r}}$, or its expectation in a weak sense.
        Moreover, we relax the energy-causality and battery-capacity constraints within each block, so the (optimal) block update number is only required to satisfy the energy-causality and (battery-capacity-induced) no-energy-overflow constraints in the block level.
        Therefore, the block update number $N_{\text{u}}$ is required to satisfy
        \begin{gather}
            N_{\text{u}}
            \le \expect N_{\text{r}}
            = n\eta \quad \text{(request-driven constraint)},\\
            N_{\text{u}}
            \le \hat{B}_i + \check{E}_i \quad \text{(energy-causality constraint)},\\
            \begin{multlined}
                \hat{B}_i + \check{E}_i - N_{\text{u}}
                \le \overline{B} - 1 \\
                \hspace{4.5em}\text{(no-energy-overflow constraint)}.
            \end{multlined}
        \end{gather}
        Equivalently, we must have
        \begin{equation}
            N_{\text{u}}
            \in [\clip{\hat{B}_i+\check{E}_i-\overline{B}+1}{0}{n\eta}, \upperclip{\hat{B}_i+\check{E}_i}{n\eta}],
        \end{equation}
        Note that the block energy arrival may not be an integer, so we have to choose the inferred battery level of the next block (instead of the current block update number) to represent the block action.
        We have
        \begin{multline}
            \hat{B}_{i+1}
            = \upperclip{\hat{B}_i+\check{E}_i-N_{\text{u}}}{\overline{B}-1} \\
            \in [\clip{\hat{B}_i+\check{E}_i-n\eta}{0}{\overline{B}-1}, \upperclip{\hat{B}_i+\check{E}_i}{\overline{B}-1}].
        \end{multline}

        \item We use an estimation of the offline-optimal total on-demand AoI within a block as the block reward.
    \end{enumerate}

    With all these ideas in mind, we define the simplified MDP as follows.

    \begin{definition}
        \label{df:smdp}
        The simplified MDP is defined as a 4-tuple $(\mathcal{Z},\mathcal{Q},p(z'|z,q),c(z,q))$, where
        \begin{itemize}
            \item $\mathcal{Z}\eqdef \integersOf{[0,\overline{B})}\times \{\hat{e}_{-1},\hat{e}_0,\hat{e}_1\}$ is the block state space and a state $z\in\mathcal{Z}$ is a pair $(\hat{b},\check{e})$ of the inferred battery level $\hat{b}$ and the block energy arrival $\check{e}$;
            \item $\mathcal{Q}\eqdef \integersOf{[0,\overline{B})}$ is the block action space, and for each state $z=(\hat{b},\check{e})$, the set of all allowable actions is
            \begin{equation}
                \mathcal{Q}_z
                \eqdef \integersOf{[\clip{\hat{b}+\check{e}-n\eta}{0}{\overline{B}-1}, \upperclip{\hat{b}+\check{e}}{\overline{B}-1}]};
            \end{equation}
            \item $p(z'|z,q) = 1\{\hat{b}'=q\} P_{\check{E}}(\check{e}')$ is the transition probability from state $z=(\hat{b},\check{e})$ to state $z'=(\hat{b}',\check{e}')$ given action $q$;
            \item $c(z,q)$ is the cost function defined by
            \begin{align}
                &c(z,q) \notag \\
                &\eqdef
                \begin{dcases}
                    n\eta - \frac{n}{2} + \frac{n^2\eta}{2\xi u}, &\frac{n}{\xi u}\le \overline{\Delta}, \\
                    \overline{\Delta}n\eta - (\overline{\Delta}-1)\xi u \\
                    \quad{} - \frac{(\overline{\Delta}-1)(\overline{\Delta}-2)\xi u(n\eta-\xi u)}{2(n-\xi u)}, &\frac{n}{\xi u} > \overline{\Delta},
                \end{dcases} \label{eq:estimation_c}
            \end{align}
            with $u\eqdef \upperclip{\hat{b}+\check{e}-q}{n\eta}$.
            \proofref[Derivation]{in Section~\ref{proof:c-value}}
        \end{itemize}
    \end{definition}

    The Bellman equation for the above MDP is
    \begin{equation}
        \check{g}^*+\check{h}(z)
        = \min_{q\in\mathcal{Q}_z} \left\{ c(z,q)+\sum_{j=-1}^1 P_{\check{E}}(\hat{e}_j)\check{h}(q,\hat{e}_j) \right\}, \label{eq:smdp_bellman}
    \end{equation}
    where $\check{g}^*$ is the optimal cost, and $\check{h}:\mathcal{Z}\to\real$ is the bias function.
    The optimal cost $g^*$ and the post-update function $\tilde{h}(\hat{b})$ of the original pBSI problem can be estimated by
    \begin{subequations}
        \begin{align}
            g^*
            &\approx \frac{\check{g}^*}{n}, \\
            \tilde{h}(\hat{b})
            &\approx \sum_{j=-1}^{1} P_{\check{E}}(\hat{e}_j)\check{h}(\hat{b},\hat{e}_j).
        \end{align}\label{eq:post_update_value_estimation}%
    \end{subequations}
    Therefore, we can apply a value iteration algorithm to solve the simplified MDP and then use~\eqref{eq:post_update_value_estimation} to obtain $g^*$ and $\tilde{h}(\hat{b})$.
    The details are given in Algorithm~\ref{alg:puvi}.

    \begin{algorithm}
        \caption{Post-Update Value Iteration Algorithm}\label{alg:puvi}
        \begin{algorithmic}[1]
            \State Initialization: $V(z)\gets 0, \check{h}(z)\gets 0,~\forall z\in\mathcal{Z}$, choose a reference state $z_{\text{ref}}$, a suitable $n$, and a small $\vartheta>0$.
            \Repeat
                \State $V'(z)\gets V(z),~\forall z\in\mathcal{Z}$
                \For{$ z\in \mathcal{Z} $}
                    \State $V(z)\gets \min_{q\in\mathcal{Q}_z} \bigl( c(z,q)$
                    \State \hspace{8em} $+ \sum_{j=-1}^1 P_{\check{E}}(\hat{e}_j)\check{h}(q,\hat{e}_j) \bigr)$
                \EndFor
                \State $\theta\gets \max_{z\in\mathcal{Z}}(V(z)-V'(z))-\min_{z\in\mathcal{Z}}(V(z)-V'(z))$
                \State $\check{h}(z)\gets V(z)-V(z_{\text{ref}}),~\forall z\in\mathcal{Z}$
            \Until{$\theta<\vartheta$}
            \State Compute $\tilde{h}(\hat{b})$ using \eqref{eq:post_update_value_estimation}
        \end{algorithmic}
    \end{algorithm}

    It is worth noting that Algorithm~\ref{alg:puvi} only computes $\tilde{h}(\hat{b})$ for integer values of $\hat{b}$, so interpolation (e.g., linear interpolation) is required for using the CN policy (Definition~\ref{df:cn_policy} and Theorem~\ref{th:n-value}).

    As an easy consequence of Definition~\ref{df:smdp} and Algorithm~\ref{alg:puvi}, we have the following proposition.

    \begin{proposition}
        \label{pr:smdp_size}
        The size of the state space $\mathcal{Z}$ is $3\overline{B}$, and the size of the action space $\mathcal{Q}$ is $\overline{B}$.
        The per-iteration time complexity of Algorithm~\ref{alg:puvi} is $\mathrm{O}(\overline{B}^2)$.
    \end{proposition}

    \begin{sketch}
        Observe that the number of multiplications in each iteration is at most $3|\mathcal{Z}||\mathcal{Q}| = 9\overline{B}^2$.
    \end{sketch}

    \section{Proofs of Main Results in Section~\ref{sec:general_pbsi}}\label{sec:proofs}

    \subsection{Proof of Theorem~\ref{th:lower_bound}}\label{proof:lower_bound}

    \begin{IEEEproof}
    [Proof of Theorem~\ref{th:lower_bound}]
        Consider the status update of a sensor node from time slot $1$ to $n$.
        In order to establish a lower bound on the expected time-average on-demand AoCSI under any policy, we assume that $\Delta_1=1$ and eBSI is available for decision-making.

        By Lemma~\ref{le:sequence_aoi}, we have
        \begin{align}
            &\expect \left( \sum_{t=1}^{n} R_t\upperclip{\Delta_{t+1}}{\overline{\Delta}} \middle| \mathcal{N} \right) \notag \\
            &\ge N_{\text{su}} + \frac{\eta(N_{\text{su}}+1)}{2} \chi_1\left(\frac{N_{\text{u}} + |\mathcal{T}_{\text{ni}}|+1}{N_{\text{su}}+1}\right) \notag \\
            &\quad + \frac{(1-\eta)(N_{\text{su}}+1)}{2} \chi_1\left(\frac{N_{\text{u}}+1}{N_{\text{su}}+1}\right) \notag \\
            &= \eta\chi_2(N_{\text{su}}+1, N_{\text{u}}+|\mathcal{T}_{\text{ni}}|+1) \notag \\
            &\quad + (1-\eta)\chi_2(N_{\text{su}}+1, N_{\text{u}}+1) - 1, \label{eq:total_aoi_lower_bound}
        \end{align}
        where $\mathcal{N}$ is defined by~\eqref{eq:sequence_aoi.knowledge}, $N_{\text{su}}$ is the number of successful updates, $N_{\text{u}}$ is the number of status-update transmission, and
        \begin{align*}
            \chi_2(x,y)
            &\eqdef x + \frac{x}{2} \chi_1\left(\frac{y}{x}\right) \\
            &=
            \begin{dcases}
                \frac{y^2}{2x}+\frac{y}{2}, &\text{$\frac{y}{x}\le \overline{\Delta}-\frac{1}{2}$}, \\
                \overline{\Delta}y - \frac{1}{2} \left(\overline{\Delta}-\frac{1}{2}\right)^2 x, &\text{$\frac{y}{x} > \overline{\Delta}-\frac{1}{2}$}.
            \end{dcases}
        \end{align*}
        Note that $\chi_2\left(xz, yz\right) = z \chi_2(x,y)$.

        Let
        \[
            r_{\text{u}}(n)
            \eqdef \frac{N_{\text{u}}+1}{n},
            \;
            r_{\text{su}}(n)
            \eqdef \frac{N_{\text{su}}+1}{N_{\text{u}}+1},
            \;
            r_{\text{req}}(n)
            \eqdef \frac{N_{\text{u}}+1}{|\mathcal{T}_{\text{ui}}|}.
        \]
        It is clear that
        \begin{subequations}
            \begin{align}
                \limsup_{n\to\infty} r_{\text{u}}(n)
                &\le \limsup_{n\to\infty} \frac{\min\{\overline{B}+\sum_{t=1}^n E_t, \sum_{t=1}^n R_t\}}{n} \notag \\
                &= \upperclip{\lambda}{\eta} \quad \text{a.s.}, \\
                \lim_{n\to\infty} r_{\text{su}}(n)
                &\eqvar{(a)} \xi \quad \text{a.s.}, \\
                \lim_{n\to\infty} r_{\text{req}}(n)
                &\eqvar{(b)} \eta \quad \text{a.s.},
            \end{align}\label{eq:r_limits}%
        \end{subequations}
        where both (a) and (b) follow from~\eqref{eq:n_su_and_n_u_freq}.

        Then from~\eqref{eq:total_aoi_lower_bound}, it follows that
        \begin{align}
            &\liminf_{n\to\infty} \frac{1}{n} \expect \left( \sum_{t=1}^{n} R_t\upperclip{\Delta_{t+1}}{\overline{\Delta}} \right) \notag \\
            &\quad = \liminf_{n\to\infty} \frac{1}{n} \expect \left( \expect \left( \sum_{t=1}^{n} R_t\upperclip{\Delta_{t+1}}{\overline{\Delta}} \middle| \mathcal{N} \right) \right) \notag \\
            &\quad \ge \liminf_{n\to\infty} \frac{1}{n} \expect \biggl[ \eta\chi_2(N_{\text{su}}+1, N_{\text{u}}+|\mathcal{T}_{\text{ni}}|+1) \notag \\
            &\quad \quad + (1-\eta)\chi_2(N_{\text{su}}+1, N_{\text{u}}+1) \biggr] \notag \\
            &\quad \gevar{(a)} \expect \liminf_{n\to\infty} \chi_3(r_{\text{su}}(n), r_{\text{u}}(n), r_{\text{req}}(n)), \label{eq:average_aoi_lower_bound.1}
        \end{align}
        where (a) follows from Fatou's lemma, and
        \[
            \chi_3(x,y,z)
            \eqdef \eta y \chi_2\left(x, \frac{1}{y} - \frac{1}{z} + 1\right) + (1-\eta) y \chi_2(x, 1).
        \]

        Suppose that $x\ge \underline{\xi}$, so $1/x\le \overline{\Delta}-1/2$.
        If $y \ge \lambda_0(z,x)$, then
        \[
            \frac{1/y - 1/z + 1}{x}
            \le \overline{\Delta}-\frac{1}{2},
        \]
        so
        \begin{multline*}
            \chi_3(x,y,z)
            = \frac{\eta}{2xy} + \eta\left(\frac{1}{2}-\frac{1-z}{xz}\right) \\
            + y\left[\frac{\eta(1-z)^2}{2xz^2} - \frac{\eta(1-z)}{2z} + \frac{1-\eta}{2x} + \frac{1-\eta}{2}\right],
        \end{multline*}
        which is strictly decreasing in $y\in [\lambda_0(z,x), y_1(x,z)]$ with
        \[
            y_1(x,z)
            \eqdef \sqrt{\frac{\eta}{\dfrac{\eta(1-z)^2}{z^2} - \dfrac{\eta x(1-z)}{z} + (1-\eta)(1+x)}}.
        \]
        Note that
        \[
            \lim_{z\to\eta} \lambda_0(z,x)
            \le \lim_{z\to\eta} z
            = \eta \quad \text{for $x\ge \underline{\eta}$}
        \]
        and
        \[
            \lim_{z\to\eta} y_1(x,z)
            = \frac{\eta}{\sqrt{1-\eta}}
            > \eta \quad \text{for $\eta\in (0,1]$}.
        \]
        If $y < \lambda_0(z,x)$, then
        \[
            \frac{1/y - 1/z + 1}{x}
            > \overline{\Delta}-\frac{1}{2},
        \]
        so
        \begin{multline*}
            \chi_3(x,y,z)
            = \eta\overline{\Delta} - y\Biggl[\eta\overline{\Delta}\left(\frac{1}{z}-1\right) \\
            + \frac{\eta}{2}\left(\overline{\Delta}-\frac{1}{2}\right)^2 x - (1-\eta) \left(\frac{1}{2x}+\frac{1}{2}\right)\Biggr],
        \end{multline*}
        which is strictly decreasing in $y\in [0, \lambda_0(z,x))$ for $z$ sufficiently close to $\eta$, because
        \begin{align*}
            &\lim_{z\to\eta} \left[\eta\overline{\Delta}\left(\frac{1}{z}-1\right) - (1-\eta) \left(\frac{1}{2x}+\frac{1}{2}\right)\right] \\
            &\quad = (1-\eta)\left(\overline{\Delta}-\frac{1}{2x}-\frac{1}{2}\right)
            \ge (1-\eta)\left(\frac{\overline{\Delta}}{2} - \frac{1}{4}\right)
            \ge 0.
        \end{align*}
        In conclusion, $\chi_3(x,y,z)$ is strictly decreasing in $y\in [0, y_1(x,z)]$ for $x\ge \xi_0$ and $z$ sufficiently close to $\eta$.

        By Proposition~\ref{pr:liminf} with~\eqref{eq:r_limits} and~\eqref{eq:average_aoi_lower_bound.1}, we finally have
        \begin{align*}
            &\liminf_{n\to\infty} \frac{1}{n} \expect \left( \sum_{t=1}^{n} R_t\upperclip{\Delta_{t+1}}{\overline{\Delta}} \right) \\
            &\quad \ge \chi_3(\xi, \upperclip{\lambda}{\eta}, \eta) \\
            &\quad =
            \begin{dcases}
                \frac{\eta}{2} + \frac{\eta}{2\xi\upperclip{\lambda}{\eta}} \\
                \quad{} - \frac{1-\eta}{\xi} \left(1 - \frac{\upperclip{\lambda}{\eta}}{2\eta}\right), &\text{$\lambda \ge \lambda_0(\eta,\xi)$}, \\
                \eta \left[\overline{\Delta} - \frac{\lambda\xi}{2}\left(\overline{\Delta}-\frac{1}{2}\right)^2\right] \\
                \quad{} - \lambda (1-\eta) \left(\overline{\Delta} - \frac{1}{2\xi} - \frac{1}{2}\right), &\text{$\lambda < \lambda_0(\eta,\xi)$}.
            \end{dcases}
        \end{align*}
    \end{IEEEproof}

    \begin{lemma}
        \label{le:sequence_aoi}
        Consider the status update of a sensor node from time slot $1$ to $n$ with $\Delta_1=1$.
        Let $\pi$ be a policy in the eBSI case, i.e., a mapping from $\mathcal{R}\times \integersOf{[0,\overline{B}]}\times \pintegers$ to $\mathcal{A}$.
        For a time slot $t$, if $B_t>0$ and $\pi(1,B_t,\Delta_t) = 1$, then the time slot $t$ is said to have an update intention (under the policy $\pi$); otherwise, it is said to have no update intention.
        Let $\mathcal{T}_{\text{ui}}$ denote the (index) set of all time slots with update intention, and thus its complement $\mathcal{T}_{\text{ni}}\eqdef \integersOf{[1,n]}\setminus \mathcal{T}_{\text{ui}}$ is the set of all time slots with no update intention.
        Suppose that there are $N_{\text{su}}$ successful updates, occurring at time slots $T_{\text{su}}(1) < T_{\text{su}}(2) < \cdots < T_{\text{su}}(N_{\text{su}})$.
        Let
        \begin{align}
            \mathcal{T}_{\text{su}}
            &\eqdef \{T_{\text{su}}(j): 1\le j\le N_{\text{su}}\}, \\
            \mathcal{T}_{\text{ni}}(i)
            &\eqdef \mathcal{T}_{\text{ni}} \cap (T_{\text{su}}(i),T_{\text{su}}(i+1)), \\
            \mathcal{T}_{\text{ui}}(i)
            &\eqdef \mathcal{T}_{\text{ui}} \cap (T_{\text{su}}(i),T_{\text{su}}(i+1)), \\
            N_{\text{fu}}(i)
            &\eqdef \sum_{t\in \mathcal{T}_{\text{ui}}(i)} R_t,
        \end{align}
        where $0\le i\le N_{\text{su}}$, $T_{\text{su}}(0)\eqdef 0$ and $T_{\text{su}}(N_{\text{su}}+1)\eqdef n+1$.
        Then for the policy $\pi$, the expected total on-demand AoCSI given
        \begin{equation}
            \mathcal{N}\eqdef ((T_{\text{su}}(i))_{i=1}^{N_\text{su}}, (\mathcal{T}_{\text{ni}}(i), \mathcal{T}_{\text{ui}}(i), N_{\text{fu}}(i))_{i=0}^{N_\text{su}}) \label{eq:sequence_aoi.knowledge}
        \end{equation}
        satisfies
        \begin{align}
            &\expect \left( \sum_{t=1}^{n} R_t\upperclip{\Delta_{t+1}}{\overline{\Delta}} \middle| \mathcal{N} \right) \notag \\
            &\quad \ge N_{\text{su}} + \frac{\eta}{2} \sum_{i=0}^{N_{\text{su}}} \chi_0(N_{\text{fu}}(i)+|\mathcal{T}_{\text{ni}}(i)|+1) \notag \\
            &\quad \quad + \frac{1-\eta}{2} \sum_{i=0}^{N_{\text{su}}} \chi_0(N_{\text{fu}}(i)+1) \label{eq:sequence_aoi.bound_1} \\
            &\quad \ge N_{\text{su}} + \frac{\eta(N_{\text{su}}+1)}{2} \chi_1\left(\frac{N_{\text{u}} + |\mathcal{T}_{\text{ni}}| + 1}{N_{\text{su}}+1}\right) \notag \\
            &\quad \quad + \frac{(1-\eta)(N_{\text{su}}+1)}{2} \chi_1\left(\frac{N_{\text{u}}+1}{N_{\text{su}}+1}\right), \label{eq:sequence_aoi.bound_2}
        \end{align}
        where $\chi_0(x)$ is defined by~\eqref{eq:chi_0}, $\chi_1(x)$ is defined by~\eqref{eq:chi_1}, and
        \begin{equation}
            N_{\text{u}}
            \eqdef N_{\text{su}}+\sum_{i=0}^{N_{\text{su}}} N_{\text{fu}}(i)
        \end{equation}
        is the number of status-update transmissions (successful or failed).
        Furthermore, if $\lim_{n\to\infty} |\mathcal{T}_{\text{ui}}| = \infty$ a.s., then
        \begin{equation}
            \lim_{n\to\infty} \frac{N_{\text{su}}}{|\mathcal{T}_{\text{ui}}|}
            = \eta\xi \quad \text{a.s.}
            \quad\text{and}\quad
            \lim_{n\to\infty} \frac{N_{\text{u}}}{|\mathcal{T}_{\text{ui}}|}
            = \eta \quad \text{a.s.} \label{eq:n_su_and_n_u_freq}
        \end{equation}
    \end{lemma}

    \begin{IEEEproof}
        Observe that
        \begin{equation}
            P(\{R_t=1\}|\mathcal{N})
            = \eta \quad \text{for all $t\in \mathcal{T}_{\text{ni}}$}, \label{eq:r_t_ni}
        \end{equation}
        because the realization of each $R_t$ over $t\in \mathcal{T}_{\text{ni}}$ has no impact on the evolution of the battery level and AoCSI.
        Therefore,
        \begin{align*}
            &\expect \left( \sum_{t=1}^{n} R_t\upperclip{\Delta_{t+1}}{\overline{\Delta}} \middle| \mathcal{N} \right) \\
            &\quad = -1 + \sum_{i=0}^{N_{\text{su}}} \left( 1 + \sum_{t=T_{\text{su}}(i)+1}^{T_{\text{su}}(i+1)-1} \expect (R_t\upperclip{\Delta_{t+1}}{\overline{\Delta}} | \mathcal{N}) \right) \\
            &\quad \eqvar{(a)} -1 + \sum_{i=0}^{N_{\text{su}}} \Biggl( 1 + \sum_{t\in \mathcal{T}_{\text{ni}}(i)} \eta \upperclip{\Delta_{t+1}}{\overline{\Delta}} \\
            &\quad \quad + \sum_{t\in \mathcal{T}_{\text{ui}}(i)} \expect (R_t\upperclip{\Delta_{t+1}}{\overline{\Delta}} | \mathcal{N}) \Biggr) \\
            &\quad \ge -1 + \sum_{i=0}^{N_{\text{su}}} \Biggl( 1 + \sum_{j\in\integersOf{[N_{\text{fu}}(i)+2,N_{\text{fu}}(i)+|\mathcal{T}_{\text{ni}}(i)|+1]}} \eta\upperclip{j}{\overline{\Delta}} \\
            &\quad \quad + \sum_{j\in\integersOf{[2,N_{\text{fu}}(i)+1]}} \upperclip{j}{\overline{\Delta}} \Biggr) \\
            &\quad \eqvar{(b)} -1 + \sum_{i=0}^{N_{\text{su}}} \Biggl( 1 + \frac{\eta}{2} \chi_0(N_{\text{fu}}(i)+|\mathcal{T}_{\text{ni}}(i)|+1) \\
            &\quad \quad + \frac{1-\eta}{2} \chi_0(N_{\text{fu}}(i)+1) \Biggr) \\
            &\quad = N_{\text{su}} + \frac{\eta}{2} \sum_{i=0}^{N_{\text{su}}} \chi_0(N_{\text{fu}}(i)+|\mathcal{T}_{\text{ni}}(i)|+1) \\
            &\quad \quad + \frac{1-\eta}{2} \sum_{i=0}^{N_{\text{su}}} \chi_0(N_{\text{fu}}(i)+1),
        \end{align*}
        where (a) follows from~\eqref{eq:r_t_ni}, and (b) from Proposition~\ref{pr:block_aoi}.
        This concludes~\eqref{eq:sequence_aoi.bound_1}, and~\eqref{eq:sequence_aoi.bound_2} follows from Proposition~\ref{pr:chi_1} and Jensen's inequality.

        From now on, we use the notation $\mathcal{T}_{\text{ui}}^{(n)}$ (in stead of $\mathcal{T}_{\text{ui}}$) to represent the set of all time slots with an update intention in the first $n$ time slots.
        Let
        \(
        \mathcal{T}_{\text{ui}}^{\infty}
        \eqdef \bigcup_{n=1}^\infty \mathcal{T}_{\text{ui}}^{(n)}
        \)
        denote the set of all time slots with an update intention from time slot $1$ to $\infty$.
        Moreover, let
        \begin{gather*}
            T_{\text{ui}}(i)
            \eqdef
            \min\left(\{t: |\integersOf{[1,t]\cap \mathcal{T}_{\text{ui}}^{\infty}}| = i\}\cup \{0\}\right)
        \end{gather*}
        denote the $i$-th time slot in $\mathcal{T}_{\text{ui}}^{\infty}$.
        Since $\lim_{n\to\infty} |\mathcal{T}_{\text{ui}}^{(n)}| = \infty$ a.s., we have $T_{\text{ui}}(i)\in \mathcal{T}_{\text{ui}}^{\infty}$ a.s.

        For each time slot $t$, we define the imaginary channel-condition random variable $\hat{H}_t\in \{0,1\}$ such that the relation between $H_t$ and $A_t$ can be expressed as $H_t=\hat{H}_tA_t$.
        It is obvious that $P\{\hat{H}_t=1\}=\xi$.
        Furthermore, consider the i.i.d.\ process $(R_t,\hat{H}_t,E_t)_{t=0}^\infty$, where $R_t$, $\hat{H}_t$, and $E_t$ are mutually independent and the time slot $0$ represents the imaginary terminal time slot in the case of the zero-probability event $|\mathcal{T}_{\text{ui}}^{\infty}| < \infty$.
        Consider the process $(R_{T_{\text{ui}}(i)},\hat{H}_{T_{\text{ui}}(i)})_{i=1}^\infty$, which is also i.i.d.\ and has the same distribution as $(R_t,\hat{H}_t)_{t=0}^\infty$.
        This is because $T_{\text{ui}}(i)$ only depends on $(R_{T_{\text{ui}}(j)},\hat{H}_{T_{\text{ui}}(j)})_{j=1}^{i-1}$ and $(E_t)_{t=1}^\infty$ (under the policy $\pi$) and $T_{\text{ui}}(i) > T_{\text{ui}}(i-1) > \cdots > T_{\text{ui}}(1)$ a.s.
        Observing that a successful (resp., failed) update happens at time slot $T_{\text{ui}}(i)$ if and only if $R_{T_{\text{ui}}(i)}=1$ and $\hat{H}_{T_{\text{ui}}(i)}=1$ (resp., $\hat{H}_{T_{\text{ui}}(i)}=0$), we immediately have~\eqref{eq:n_su_and_n_u_freq}.
    \end{IEEEproof}

    \begin{proposition}
        \label{pr:block_aoi}
        Consider the status update of a sensor node from time slot $1$ to $\ell$.
        If there is only one successful update at the first time slot and $P\{R_t=1\}=\hat{\eta}$ for $2\le t\le \ell$, then the expected total on-demand AoCSI is
        \begin{equation}
            \expect\sum_{t=1}^{\ell} R_t\upperclip{\Delta_{t+1}}{\overline{\Delta}}
            = 1 + \hat{\eta} \sum_{t=2}^{\ell} \upperclip{t}{\overline{\Delta}}
            = 1 + \frac{\hat{\eta}}{2} \chi_0(\ell),
        \end{equation}
        where
        \begin{align}
            \chi_0(x)
            &\geqdef 2x\upperclip{x}{\overline{\Delta}} - \upperclip{x}{\overline{\Delta}}^2 + \upperclip{x}{\overline{\Delta}} - 2 \notag \\
            &=
            \begin{cases}
                x^2+x-2, &\text{$x\le\overline{\Delta}$}, \\
                2\overline{\Delta}x - \overline{\Delta}^2 + \overline{\Delta} - 2, &\text{$x>\overline{\Delta}$}.
            \end{cases} \label{eq:chi_0}
        \end{align}
    \end{proposition}

    Note that $R_2,\ldots,R_{\ell}$ are not required to be mutually independent in Proposition~\ref{pr:block_aoi}.

    \begin{IEEEproof}
        Since there is only one successful update at the first time slot, we have $R_1=1$, and $\Delta_{t+1}=t$ for $1\le t\le \ell$.
        Therefore, the expected total on-demand AoCSI is
        \begin{align*}
            &\expect\sum_{t=1}^{\ell} R_t\upperclip{\Delta_{t+1}}{\overline{\Delta}} \\
            &\quad = \mathbb{E} \left(1 + \sum_{t=2}^{\ell} R_t \upperclip{t}{\overline{\Delta}} \right)
            = 1 + \hat{\eta} \sum_{t=2}^{\ell} \upperclip{t}{\overline{\Delta}} \\
            &\quad = 1 - \hat{\eta} + \hat{\eta} \sum_{t=1}^{\ell} \upperclip{t}{\overline{\Delta}}
            = 1 - \hat{\eta} + \hat{\eta} \sum_{t=1}^{\ell} \sum_{j=1}^{t} 1\{j\le \overline{\Delta}\} \\
            &\quad = 1 - \hat{\eta} + \hat{\eta} \sum_{j=1}^{\ell} 1\{j\le \overline{\Delta}\} \sum_{t=j}^{\ell} 1 \\
            &\quad = 1 - \hat{\eta} + \hat{\eta} \sum_{j=1}^{\upperclip{\ell}{\overline{\Delta}}} (\ell-j+1) \\
            &\quad = 1 - \hat{\eta} + \hat{\eta} \upperclip{\ell}{\overline{\Delta}} \frac{2\ell-\upperclip{\ell}{\overline{\Delta}}+1}{2} \\
            &\quad = 1 + \frac{\hat{\eta}}{2} (2\ell\upperclip{\ell}{\overline{\Delta}} - \upperclip{\ell}{\overline{\Delta}}^2 + \upperclip{\ell}{\overline{\Delta}} - 2).
        \end{align*}
    \end{IEEEproof}

    \begin{proposition}
        \label{pr:chi_1}
        The function $\chi_0(x)$ defined by~\eqref{eq:chi_0} is bounded below by
        \begin{equation}
            \chi_1(x)
            \geqdef
            \begin{dcases}
                x^2+x-2, &\text{$x\le\overline{\Delta}-\frac{1}{2}$}, \\
                2\overline{\Delta}x - \overline{\Delta}^2 + \overline{\Delta} - \frac{9}{4}, &\text{$x>\overline{\Delta}-\frac{1}{2}$},
            \end{dcases} \label{eq:chi_1}
        \end{equation}
        which is continuously differentiable, strictly increasing, and convex on $[0,+\infty)$.
    \end{proposition}

    \begin{sketch}
        It suffices to show that $\chi_1(x)$ is continuously differentiable, and other facts are straightforward.
        Let $\overline{\Delta}'\eqdef \overline{\Delta}-\frac{1}{2}$.
        It is easy to verify that $\chi_1(\overline{\Delta}') = \overline{\Delta}^2 - \frac{9}{4} = \lim_{x\downarrow \overline{\Delta}'} \chi_1(x)$.
        Finally, observe that the left and right derivatives of $\chi_1(x)$ at $x=\overline{\Delta}'$ are both $2\overline{\Delta}$.
    \end{sketch}

    \begin{proposition}
        \label{pr:liminf}
        Let $f(x,y)$ be a continuous mapping from $\preal \times \preal^k$ to $\nnreal$.
        Let $(x_i)_{i=1}^\infty$ be a sequence of positive real numbers and $(y_i)_{i=1}^\infty=(y_{i,1},y_{i,2},\ldots,y_{i,k})_{i=1}^\infty$ be a sequence of $k$-dimensional positive real vectors.
        Suppose that $\limsup_{i\to\infty} x_i \le x_0$ and $\lim_{i\to\infty} y_i = y_0$, where $x_0$ and $y_0$ are both positive.
        If $f(x,y)$ is decreasing in $x\in [0,g(y)]$ for all fixed $y$ in a neighborhood $\mathcal{N}(y_0)$ of $y_0$, where $x_0 < g(y_0)$ and $g(y)$ is continuous at $y_0$, then
        \begin{equation}
            \liminf_{i\to\infty} f(x_i,y_i)
            \ge f(x_0, y_0).
        \end{equation}
    \end{proposition}

    \begin{IEEEproof}
        Let $\epsilon_0\eqdef (\upperclip{g(y_0)}{x_0+1}-x_0) / 2$.
        For any $\epsilon \in (0,\epsilon_0)$, there exists $i_0\ge 1$ such that
        \[
            x_i
            < x_0 + \epsilon
            < \upperclip{g(y_0)}{x_0+1} - \epsilon
            < g(y_i)
        \]
        for all $i\ge i_0$.
        Then for $i\ge i_0$,
        \[
            \liminf_{i\to\infty} f(x_i,y_i)
            \ge \liminf_{i\to\infty} f(x_0+\epsilon, y_i)
            = f(x_0+\epsilon, y_0),
        \]
        and hence
        \[
            \liminf_{i\to\infty} f(x_i,y_i)
            \ge \sup_{\epsilon\in (0,\epsilon_0)} f(x_0+\epsilon, y_0)
            = f(x_0, y_0).
        \]
    \end{IEEEproof}

    \subsection{Proofs of Theorem~\ref{th:n-value} and Propositions~\ref{pr:kth_request_estimation} and~\ref{pr:xth_step_estimation}}\label{proof:n-value}

    \begin{IEEEproof}
    [Proof of Theorem~\ref{th:n-value}]
        It is clear that the initial expected sensor battery level is approximately $b_0 = \upperclip{\hat{b}+\lambda \delta}{\overline{B}}$.
        Since $\bar{n}\le b_0$, the ($N$-step) total value under the strategy $\Gamma_1$ can be estimated by
        \[
            \sum_{i=1}^{\bar{n}} \xi(1-\xi)^{i-1} \hat{V}(b_0,\delta,i) + (1-\xi)^{\bar{n}} W_N,
        \]
        where $\hat{V}(b_0,\delta,i)$ is the total value under $\Gamma_1$ given that the first successful update occurs at the $i$-th time slot with request (Proposition~\ref{pr:kth_request_estimation}), and $W_N$ is the total value under $\Gamma_1$ given that the first $\bar{n}$ updates all fail.

        Let $\hat{T}_i$ denote the (one-based) step index of the $i$-th time slot with request, counted from the current time slot.
        Then, $W_N$ can be estimated by
        \begin{align}
            &W_N \notag \\
            &\approx \expect \left( \sum_{i=1}^{\bar{n}} \upperclip{\delta+\hat{T}_i}{\overline{\Delta}} + \tilde{V}_{N-\hat{T}_{\bar{n}}}(b_0-\bar{n}+\lambda \hat{T}_{\bar{n}}, \delta+\hat{T}_{\bar{n}}, X)\right) \notag \\
            &\approx \sum_{i=1}^{\bar{n}} \upperclip{\delta+\expect\hat{T}_i}{\overline{\Delta}} + \tilde{V}_{N-\expect\hat{T}_{\bar{n}}}(b_0-\bar{n}+\lambda \expect\hat{T}_{\bar{n}}, \delta+\expect\hat{T}_{\bar{n}}, \expect X) \notag \\
            &\eqvar{(a)} \sum_{i=1}^{\bar{n}} \upperclip{\delta+\phi_0(i)}{\overline{\Delta}} \notag \\
            &\quad + \tilde{V}_{N-\phi_0(\bar{n})}(b_0-\bar{n}+\lambda\phi_0(\bar{n}), \delta+\phi_0(\bar{n}), \expect X) \notag \\
            &= \psi_1(\delta,\bar{n}) + \tilde{V}_{N-\phi_0(\bar{n})}(b_1, \psi_0(\delta,\bar{n}), \expect X), \label{eq:n-value-proof.eq1}
        \end{align}
        where $\tilde{V}_{N}(b,\delta,x)$ is the total value under $\Gamma_1$ given that the first successful update occurs at the $x$-th step (Proposition~\ref{pr:xth_step_estimation}), $X\eqdef Y-\hat{T}_{\bar{n}}$ with $Y$ being the step index (counted from the current time slot) of the first successful update, and (a) follows from Lemma~\ref{le:expected_request_time}.

        Next, we estimate $\expect X$.
        After the update failure at the step $\hat{T}_{\bar{n}}$ (the $\bar{n}$-th time slot with request), we still need $1/\xi$ updates on average for a successful update.
        Then, the expected value $x\eqdef \expect X$ satisfies the following equation approximately:
        \[
            \min\left\{b+\lambda x, 1+\eta(x-1)\right\}
            \approx \frac{1}{\xi},
        \]
        where $b$ is the expected sensor battery level after the step $\hat{T}_{\bar{n}}$.
        Solving this equation, we obtain
        \[
            x
            = \max\left\{\frac{1}{\xi\lambda} - \frac{b}{\lambda}, \frac{1}{\xi\eta}-\frac{1}{\eta}+1\right\}.
        \]
        This combined with \eqref{eq:n-value-proof.eq1} gives \eqref{eq:n-value.1}.

        Eq.~\eqref{eq:n-value.2} is an easy sequence of~\eqref{eq:n-value.1} because $\Gamma_2$ can be equivalently regarded as $\Gamma_1$ with one additional energy unit but failing at the first time slot with request.
        Note that a slight adjustment (see Eqs.~\eqref{eq:n-value.2}, \eqref{eq:b_0_adjustment}, and~\eqref{eq:b_2}) of the initial expected battery level $b_0$ is required to take into account the case where the harvested energy during the first $1/\eta$ time slots cannot be charged to the battery due to a close-to-capacity initial battery level.
    \end{IEEEproof}

    \begin{lemma}
        \label{le:expected_request_time}
        Let $\hat{T}_i$ be the one-based step index of the $i$-th time slot with request, counted from the current time slot.
        Then, $\expect \hat{T}_i = \phi_0(i)$, where $\phi_0(i)$ is defined by~\eqref{eq:phi_0}.
    \end{lemma}

    \begin{IEEEproof}
        Let $\Delta\hat{T}_i\eqdef \hat{T}_{i}-\hat{T}_{i-1}$ for $i\ge 2$.
        It is clear that $\expect \Delta\hat{T}_i = 1 / \eta$.
        Therefore,
        \[
            \expect \hat{T}_i
            = \expect \left( \hat{T}_1 + \sum_{j=2}^i \Delta\hat{T}_i \right)
            = 1 + \frac{i-1}{\eta}
            = \phi_0(i).
        \]
    \end{IEEEproof}

    \begin{IEEEproof}
    [Proof of Proposition~\ref{pr:kth_request_estimation}]
        Since the first successful update occurs at the $i$-th time slot with request, the $N$-step total value under $\Gamma_1$ can be estimated by
        \begin{align*}
            \hat{V}_N(b,\delta,i)
            &\approx \expect \Biggl( \sum_{j=1}^{i-1} \upperclip{\delta+\hat{T}_j}{\overline{\Delta}} + 1 + (N-\hat{T}_i)g^* \\
            &\quad + \tilde{h}(b-i+\lambda(\hat{T}_i-1)) \Biggr) \\
            &\approx \sum_{j=1}^{i-1} \upperclip{\delta+\expect\hat{T}_j}{\overline{\Delta}} + 1 + (N-\expect\hat{T}_i)g^*\\
            &\quad + \tilde{h}(b-i+\lambda\expect\hat{T}_i-\lambda) \\
            &\eqvar{(a)} \sum_{j=1}^{i-1} \upperclip{\delta+\phi_0(j)}{\overline{\Delta}} + 1 + (N-\phi_0(i))g^* \\
            &\quad + \tilde{h}(b-i+\lambda\phi_0(i)-\lambda) \\
            &= 1 + \psi_1(\delta,i-1) + (N-\phi_0(i))g^* \\
            &\quad + \tilde{h}(b-i+\lambda\phi_0(i)-\lambda),
        \end{align*}
        where $\hat{T}_i$ denotes the one-based step index (counted from the current time slot) of the $i$-th time slot with request, and (a) follows from Lemma~\ref{le:expected_request_time}.
    \end{IEEEproof}

    \begin{IEEEproof}
    [Proof of Proposition~\ref{pr:xth_step_estimation}]
        We first suppose that $x$ is a positive integer.
        Since the first successful update occurs at the $x$-th step, the $N$-step total value under $\Gamma_1$ can be estimated by
        \begin{align*}
            \tilde{V}_N(b,\delta,i)
            &\approx \expect \Biggl[ \sum_{j=1}^{x-1} R_j \upperclip{\delta+j}{\overline{\Delta}} + 1 + (N-x)g^* \\
            &\quad + \tilde{h}\Biggl(\lowerclip*{b - 1 - \sum_{j=1}^{x-1} R_j + \lambda (x-1)}{0}\Biggr) \Biggr] \\
            &\approx \eta \sum_{j=1}^{x-1} \upperclip{\delta+j}{\overline{\Delta}} + 1 + (N-x)g^* \\
            &\quad + \tilde{h}(\lowerclip{b-(1+\eta(x-1))+\lambda x-\lambda}{0}) \\
            &= 1 + \eta \psi_1(\delta,x-1) + (N-x)g^* \\
            &\quad + \tilde{h}(\lowerclip{b-\phi_2(x)+\lambda x-\lambda}{0}),
        \end{align*}
        where $\phi_2(x)$ is defined by~\eqref{eq:phi_2}.

        As for a general real number $x\ge 1$, we consider an interpolation of the above estimation:
        \[
            (1-t) \tilde{V}_N(b,\delta,\lfloor x \rfloor) + t \tilde{V}_N(b,\delta,\lceil x \rceil),
        \]
        where $t=x-\lfloor x \rfloor$.
        Therefore, we have
        \begin{align*}
            \tilde{V}_N(b,\delta,x)
            &\approx 1 + \eta\psi_2(\delta,x-1) + (N-x)g^* \\
            &\quad + (1-t)\tilde{h}(\lowerclip{b-\phi_1(\lfloor x \rfloor)+\lambda x}{0}) \\
            &\quad + t\tilde{h}(\lowerclip{b-\phi_1(\lceil x \rceil)+\lambda x}{0}) \\
            &\approx 1 + \eta\psi_2(\delta,x-1) + (N-x)g^* \\
            &\quad + \tilde{h}(\lowerclip{b-\phi_1(x)+\lambda x}{0}),
        \end{align*}
        where $\psi_2(b,\delta)$ is defined by~\eqref{eq:psi_2}.
    \end{IEEEproof}

    \subsection{Derivation of Eq.~\eqref{eq:estimation_c}}\label{proof:c-value}

    \begin{IEEEproof}[Derivation of Eq.~\eqref{eq:estimation_c}]
        Let $N_{\text{r}}$ be the number of requests in a block (of $n$ time slots).
        Let $N_{\text{u}}$ be the number of updates in the block.
        We suppose that $\expect(N_{\text{u}}|N_{\text{r}})\approx u \eqdef \upperclip{\hat{b}+\check{e}-q}{n\eta}$ under the optimal offline policy.
        Let $N_{\text{su}}$ be the number of successful updates in the block.
        It is clear that $\expect(N_{\text{su}}|N_{\text{r}}) = \xi \expect (N_{\text{u}}|N_{\text{r}}) \approx \xi u$.

        Given $N_{\text{r}}$ and $N_{\text{su}}$, the block can be divided into $N_{\text{su}}$ segments, each ending with a successful update.
        Let $T_{\text{su}}(j)$ denote the time slot with the $j$-th successful update and let $\mathcal{T}_{\text{su}}$ be the set of all time slots with a successful update.
        Observe that the probabilities $P\{R_t=1\}$ for all time slots with no successful update are approximately equal, so we have
        \begin{align*}
            P(\{R_t=1\}|N_{\text{r}},N_{\text{su}})
            &\approx \frac{\sum_{t\notin \mathcal{T}_{\text{su}}} P(\{R_t=1\}|N_{\text{r}},N_{\text{su}})}{n-N_{\text{su}}} \\
            &= \frac{\expect (\sum_{t\notin \mathcal{T}_{\text{su}}} 1\{R_t=1\}|N_{\text{r}},N_{\text{su}})}{n-N_{\text{su}}} \\
            &= \hat{\eta}
            \eqdef \frac{N_{\text{r}}-N_{\text{su}}}{n-N_{\text{su}}}
        \end{align*}
        for all time slot $t$ with no successful update.

        From Proposition~\ref{pr:block_aoi}, it follows that the expected total on-demand AoCSI over the block given $N_{\text{r}}$ and $N_{\text{su}}$ is
        \begin{align*}
            &\expect \left(\sum_{t=1}^{n} R_t\upperclip{\Delta_{t+1}}{\overline{\Delta}} \middle| N_{\text{r}},N_{\text{su}} \right) \\
            &= N_{\text{su}} + \frac{\hat{\eta}}{2} \expect \left( \sum_{j=1}^{N_{\text{su}}} \chi_0(T_{\text{su}}(j)-T_{\text{su}}(j-1)) \middle| N_{\text{r}},N_{\text{su}} \right) \\
            &\relvar{(a)}{\gtrapprox} N_{\text{su}} + \frac{\hat{\eta}N_{\text{su}}}{2} \chi_0\left(\frac{n}{N_{\text{su}}}\right)
            = \hat{\chi}(N_{\text{su}},N_{\text{r}}) \\
            &\eqdef
            \begin{dcases}
                N_{\text{r}} - \frac{n}{2} + \frac{nN_{\text{r}}}{2N_{\text{su}}}, & \frac{n}{N_{\text{su}}}\le \overline{\Delta}, \\
                \overline{\Delta}N_{\text{r}} - (\overline{\Delta}-1)N_{\text{su}} \\
                \quad {} - \frac{(\overline{\Delta}-1)(\overline{\Delta}-2)N_{\text{su}}(N_{\text{r}}-N_{\text{su}})}{2(n-N_{\text{su}})}, &\frac{n}{N_{\text{su}}} > \overline{\Delta},
            \end{dcases}
        \end{align*}
        where $T_{\text{su}}(0)$ denotes the time slot before the block, and (a) follows from Jensen's inequality with Proposition~\ref{pr:chi_1} approximately.
        The lower bound $\hat{\chi}(N_{\text{su}},N_{\text{r}})$ serves as a good estimation for the offline-optimal total on-demand AoCSI over the block given $N_{\text{r}}$ and $N_{\text{su}}$.
        It is decreasing and approximately convex in $N_{\text{su}}$ for fixed $N_{\text{r}}$.
        Therefore, the offline-optimal expected total on-demand AoCSI over the block can be estimated by
        \begin{align*}
            &\expect \sum_{t=1}^{n} R_t\upperclip{\Delta_{t+1}}{\overline{\Delta}} \\
            &\quad \ge \expect \left( \expect \left( \hat{\chi}(N_{\text{su}},N_{\text{r}}) \middle| N_{\text{r}} \right) \right) \\
            &\quad \relvar{(a)}{\gtrapprox} \expect \hat{\chi}(\expect(N_{\text{su}}|N_{\text{r}}),N_{\text{r}})
            \approx \expect \hat{\chi}(\xi u,N_{\text{r}})
            = \hat{\chi}(\xi u,n\eta) \\
            &\quad =
            \begin{dcases}
                n\eta - \frac{n}{2} + \frac{n^2\eta}{2\xi u}, &\frac{n}{\xi u}\le \overline{\Delta}, \\
                \overline{\Delta}n\eta - (\overline{\Delta}-1)\xi u \\
                \quad {} - \frac{(\overline{\Delta}-1)(\overline{\Delta}-2)\xi u(n\eta-\xi u)}{2(n-\xi u)}, &\frac{n}{\xi u} > \overline{\Delta},
            \end{dcases}
        \end{align*}
        where (a) follows from Jensen's inequality for conditional expectation approximately.
    \end{IEEEproof}

    \section{Weighted Update Gain Competition: The Multi-Sensor Case}\label{sec:wugc}

    By proposing the CN policy, we have successfully solved the single-sensor pBSI problem.
    In the case of multiple sensors with the simultaneous update constraint~\eqref{eq:update-constraint}, we first use the CN policy to determine all the sensors that may be updated.
    If the number of sensors to be updated is less than or equal to $K_0$, then all these sensors are commanded to update.
    Otherwise, we need to choose $K_0$ sensors to minimize the long-term average cost.

    We take a heuristic approach to solve this problem by simply considering the instantaneous cost difference between updating and not updating.
    The \term{weighted update gain} of each sensor $k$ is defined as
    \begin{align}
        &\mathrm{WUG}_k \notag \\
        &\eqdef \alpha_k \left[ \upperclip{\Delta_{t,k}+1}{\overline{\Delta}_k} - \left(\omega_k+(1-\omega_k) \upperclip{\Delta_{t,k}+1}{\overline{\Delta}_k} \right) \right] \notag \\
        &= \alpha_k \omega_k (\upperclip{\Delta_{t,k}+1}{\overline{\Delta}_k} - 1), \label{eq:wug}
    \end{align}
    where
    \begin{equation}
        \omega_k
        \approx \xi_k \{1\{\hat{B}_{t,k}>0\}+[1-(1-p_{1,k})^{D_{t,k}}]1\{\hat{B}_{t,k}=0\}\}
    \end{equation}
    is the probability that sensor $k$ can successfully update, and $p_{1,k}\eqdef P\{E_{1,k}>0\}$.
    Note here that we use the AoFBL $D_{t,k}$ to estimate the non-empty-battery probability of sensor $k$.
    Then, we can choose the top $K_0$ sensors with the highest weighted update gains to update.
    Algorithm~\ref{alg:wugc} summarizes the weighted update gain competition (WUGC) algorithm.

    \begin{algorithm}
        \caption{The CN Policy with the Weighted Update Gain Competition (WUGC-CN)}
        \label{alg:wugc}
        \begin{algorithmic}[1]
            \For{each time slot $t$}
                \State determine the set $\kappa_t$ of sensors to be updated using the CN policy.
                \If{$|\kappa_t|\le K_0$}
                    \State $A_{t,k} \gets 1\{k\in\kappa_t\}$ for all $k\in\mathcal{K}$
                \Else
                    \State $\kappa_t' \gets \arg \max_{\iota\subseteq \kappa_t, |\iota|=K_0} \sum_{i\in \iota} \mathrm{WUG}_i$
                    \State $A_{t,k} \gets 1\{k\in\kappa_t'\}$ for all $k\in\mathcal{K}$
                \EndIf
            \EndFor
        \end{algorithmic}
    \end{algorithm}

    \section{Numerical Results}\label{sec:simulation}

    In this section, we analyze the proposed policies and evaluate their performance through numerical methods.
    The general settings for all numerical results are as follows:
    (i) the battery capacity $\overline{B}_k$ (for each sensor $k$) is set to $15$ units of energy;
    (ii) the maximum admissible AoCSI $\overline{\Delta}_k$ is set to $48$;
    (iii) the weight $\alpha_k$ is set to $1$;
    (iv) the block length $n_k$ of the simplified MDP for estimating the post-update value function is set to $\overline{B}_k/\lambda_k$;
    and (v) one thousand episodes, each of length $10^4$, are generated for evaluating the simulation performance in each case.

    \subsection{Structural Properties of the CN Policy}\label{subsec:cn_structure}

    We analyze the structural properties of the CN policy (Definition~\ref{df:cn_policy}, Theorem~\ref{th:n-value}, and Algorithm~\ref{alg:puvi}) for a single sensor under the Bernoulli energy arrival distribution
    \begin{equation}
        P_{E_t}(e)
        \eqdef \lambda 1\{e=1\} + (1-\lambda) 1\{e=0\}. \label{eq:bernoulli}
    \end{equation}

    \begin{figure*}[htbp]
        \centering
        \includegraphics{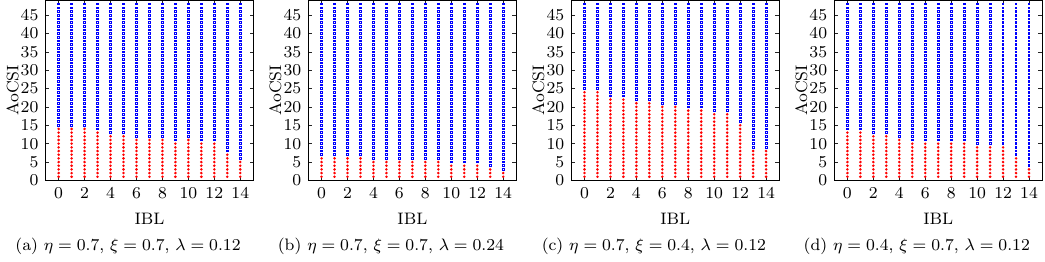}
        \caption{The structural properties of the CN policy for Bernoulli energy arrivals.}
        \label{fig:cn_policy_structure}
    \end{figure*}

    Fig.~\ref{fig:cn_policy_structure} (a) depicts the structure of the CN policy for $\eta=0.7$, $\xi=0.7$, and $\lambda=0.12$.
    Each point in the figure represents a feasible state $(\hat{b},\delta)$ of the system, a pair of the inferred battery level (IBL) and the AoCSI.
    Different markers represent different actions given by the CN policy, where the blue square indicates $a=1$ and the red circle indicates $a=0$.
    Note that the CN policy exhibits a threshold-based structure, meaning that the action is $a=1$ only when the AoCSI exceeds an IBL-dependent threshold.
    The (AoCSI) threshold decreases as the IBL increases.

    By adjusting the values of $\lambda$, $\xi$, and $\eta$, we study the impact of these parameters on the thresholds.
    Comparing Fig.~\ref{fig:cn_policy_structure} (a) with Fig.~\ref{fig:cn_policy_structure} (b), we observe that the thresholds decrease as $\lambda$ increases from $0.12$ to $0.24$.
    Comparing Fig.~\ref{fig:cn_policy_structure} (a) with Fig.~\ref{fig:cn_policy_structure} (c), we observe that the thresholds increase as $\xi$ decreases from $0.7$ to $0.4$.
    Finally, comparing Fig.~\ref{fig:cn_policy_structure} (a) with Fig.~\ref{fig:cn_policy_structure} (d), we observe that the thresholds decrease as $\eta$ decreases from $0.7$ to $0.4$.
    However, compared to other factors, the request probability $\eta$ has a smaller impact on the thresholds, so in the next subsection, we focus more on the effects of $\lambda$ and $\xi$.

    \subsection{Performance of the CN Policy for a Single Sensor}\label{subsec:cn_performance}

    We evaluate the performance of the CN policy for a single sensor under the Bernoulli and Poisson energy arrival distributions given by~\eqref{eq:bernoulli} and
    \begin{equation}
        P_{E_t}(e)
        \eqdef \frac{\lambda^e\exp(-\lambda)}{e!},
    \end{equation}
    respectively.
    Two performance metrics are introduced to assess the effectiveness of a policy.

    \begin{definition}
        The \term{additive gap} $G_+(\pi)$ and the \term{multiplicative gap} $G_\times(\pi)$ of a policy $\pi$ are defined respectively as
        \begin{equation}
            G_+(\pi)
            \eqdef g(\pi)-\Theta(\lambda,\eta,\xi)
        \end{equation}
        and
        \begin{equation}
            G_\times(\pi)
            \eqdef \frac{g(\pi)}{\Theta(\lambda,\eta,\xi)}-1,
        \end{equation}
        where $\Theta(\lambda,\eta,\xi)$ is the lower bound given in Theorem~\ref{th:lower_bound}.
    \end{definition}

    \begin{figure*}[htbp]
        \centering
        \includegraphics{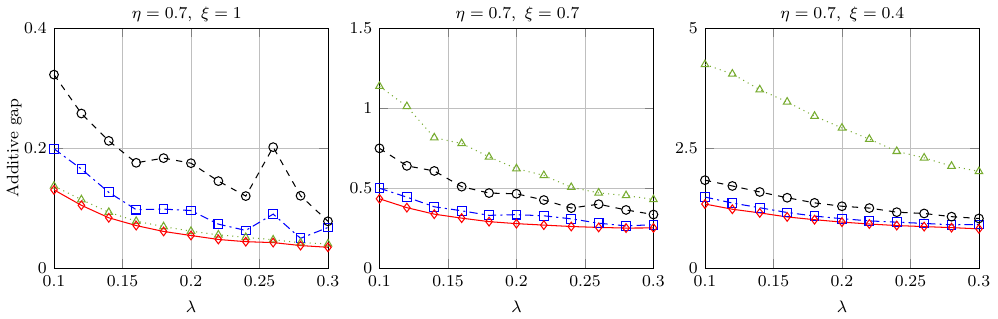}
        \includegraphics{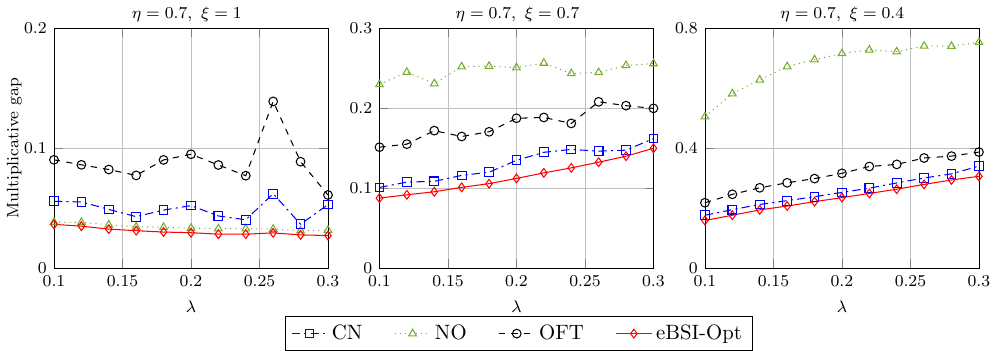}
        \caption{The addtivie and multiplicative gaps of the CN, NO, OFT, and eBSI-Opt policies for $\eta=0.7$, $\xi=1,0.7,0.4$, and Bernoulli energy arrivals with $\lambda \in \{ 0.1,0.12,\dots, 0.3\}$.}
        \label{fig:perf-bernoulli-r7-add-mul}
    \end{figure*}
    \begin{figure*}[htbp]
        \centering
        \includegraphics{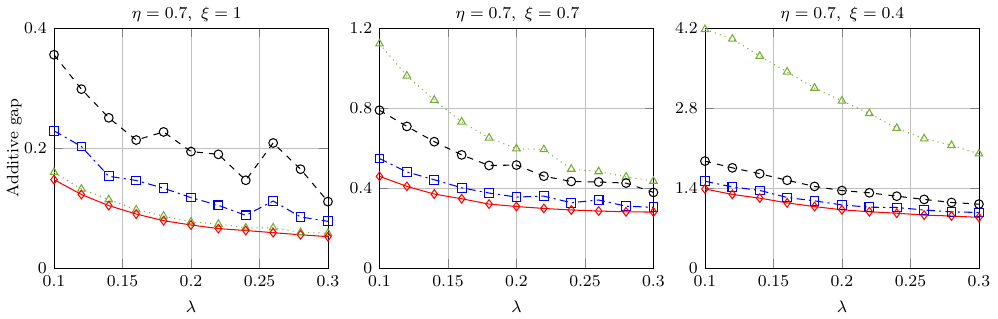}
        \includegraphics{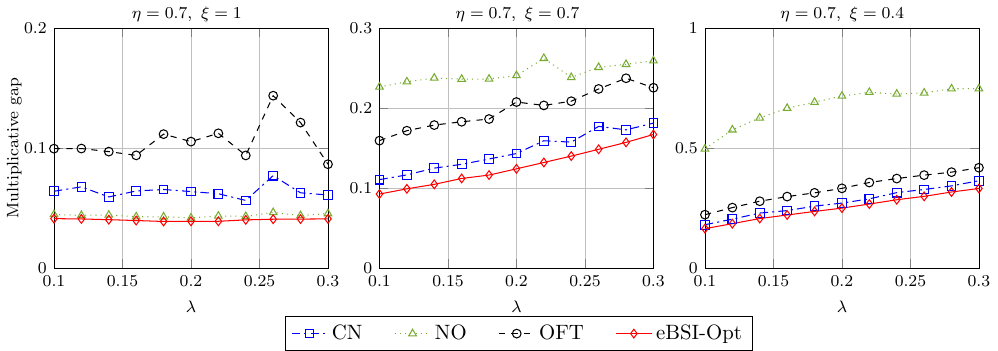}
        \caption{The additive and multiplicative gaps of the CN, NO, OFT, and eBSI-Opt policies for $\eta=0.7$, $\xi=1,0.7,0.3$, and Poisson energy arrivals with $\lambda \in \{0.1,0.12,\dots, 0.3 \}$.}
        \label{fig:perf-poisson-r7-add-mul}
    \end{figure*}

    In addition to the CN policy, we also evaluate the performance of the noiseless pBSI optimal (NO) policy with the maximum admissible AoFBL $\overline{D}=48$, the eBSI optimal (eBSI-Opt) policy~\cite[Alg.~1]{hatami2022}, and the optimal fixed-threshold (OFT) policy for comparison.
    The OFT policy employs two AoCSI updating thresholds, determined through exhaustive search, which are applied to time slots with and without requests, respectively.

    Fig.~\ref{fig:perf-bernoulli-r7-add-mul} plot the additive and multiplicative gaps of the four policies under Bernoulli energy arrival distributions.
    First, the NO policy, which is optimal for $\xi=1$, shows significant performance degradation as $\xi$ decreases from $1$ to $0.4$.
    Second, the performance gap between the CN and eBSI-Opt policies remains consistently small (averaging less than $2\%$), suggesting that the CN policy is near optimal.
    Note that the performance of the eBSI-Opt policy can serve as a lower bound for the performance of the optimal pBSI policy.
    Third, the CN policy consistently outperforms the OFT policy, demonstrating that pBSI does improve the timeliness of status updates in IoT systems.
    These three facts can also be observed in Fig.~\ref{fig:perf-poisson-r7-add-mul}.

    \subsection{Performance of the WUGC-CN Policy for Multiple Sensors}\label{sec:wugc_cn_performance}

    \begin{figure*}[htbp]
        \centering
        \includegraphics{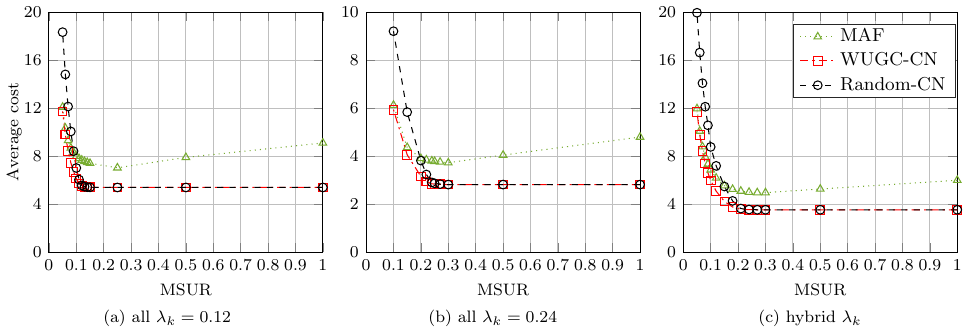}
        \caption{Performance comparison of the WUGC-CN, MAF, and random-CN policies in a $100$-sensor IoT system under Bernoulli energy arrival processes with various settings of $\lambda_k$ ($\eta_k=0.7$, $\xi_k=1,0.7,0.4$).}
        \label{fig:multi-sensor}
    \end{figure*}

    We consider an IoT system with $K=100$ sensors, where the probability of requesting each physical quantity $\zeta_k$ is $\eta_k=0.7$ for all $k \in {1, \dots, 100}$.
    Additionally, the transmission success probability $\xi_k$ is set to 1 for 33 randomly selected sensors, 0.7 for another 33 randomly selected sensors, and 0.4 for the remaining 34 sensors.
    Three experiments are conducted to evaluate the performance of the WUGC-CN policy under Bernoulli energy arrival processes.
    The compared policies include:
    \begin{itemize}
        \item The MAF policy~\cite{bedewy2021optimal}: The edge node selects $K_0$ sensors with the highest AoCSI $\Delta_{t,k}$ to command in each time slot $t$.
        \item The CN policy with random scheduling (random-CN): Different from Algorithm~\ref{alg:wugc}, the edge node randomly commands $K_0$ sensor in $\kappa_t$ to update if \( |\kappa_t| > K_0 \).
    \end{itemize}

    Fig.~\ref{fig:multi-sensor} illustrates the performance of the WUGC-CN policy for different MSURs.
    The clipped mean of the energy arrival process for each sensor in Fig.~\ref{fig:multi-sensor} (a) and (b) is set to $\lambda_k=0.12$ and $\lambda_k=0.24$, respectively.
    In Fig.~\ref{fig:multi-sensor} (c), we consider the case of hybrid energy arrival rates.
    The total 100 sensors are randomly partitioned into 4 equal-size subsets with the energy arrival rates $\lambda_k=0.12$, $0.18$, $0.24$, and $0.30$, respectively.

    At high MSUR values, the performance of WUGC-CN and random-CN is almost indistinguishable from that without the simultaneous-update constraint.
    The turning point occurs at
    \begin{equation}
        \mathrm{MSUR}
        = \bar{\lambda}
        \eqdef \frac{1}{K} \sum_{k=1}^{K} \lambda_k.
    \end{equation}
    The performance of both methods declines when $\mathrm{MSUR}<\bar{\lambda}$.
    While the performance of random scheduling significantly degrades in the low-MSUR regime, the WUGC approach remains consistently effective across all MSUR values. It is important to note that when $\mathrm{MSUR}<\bar{\lambda}$, the radio-resource constraint, rather than the energy-supply constraint, becomes the primary bottleneck for system performance, which falls outside the scope of this paper. Despite this, the performance of MAF, which focuses on the radio-resource constraint, is clearly worse than WUGC-CN across all MSUR values. This suggests that pBSI is beneficial for maintaining information timeliness in IoT systems, even when the radio-resource constraint is the dominant performance bottleneck.
    Although only the MAF policy is compared in the experiments, the WUGC-CN policy is expected to outperform all BSI-agnostic policies for MSUR values ranging from moderately below
    $\bar{\lambda}$ to $1$.
    This is because, as shown in Section~\ref{subsec:cn_performance}, the CN-policy outperforms the OFT policy, the best BSI-agnostic stationary deterministic policy, in the single-sensor case.

    \section{Conclusion}\label{sec:conclusion}

    We have investigated the status update control problem with pBSI and developed a low-complexity near-optimal policy called WUGC-CN for multi-sensor IoT systems. To the best of our knowledge, this is the first low-complexity solution for status update control with pBSI and general channels. Numerical results show a significant performance gain of pBSI compared to the optimal policy (OFT) for the nBSI case. We conclude that pBSI is beneficial for improving the timeliness of information in IoT systems and can be effectively utilized with low computational cost.

    Furthermore, the principles underlying our approach can be applied to general AoI-optimization problems in more complex scenarios, and may be integrated with universal methods such as reinforcement learning.
    By analyzing post-update states and evaluating trade-offs between variable-length actions (e.g., current vs.\ next), we successfully eliminate the computational complexity dependence on the AoI dimension.
    Additionally, our proposed online-offline hybrid approximation method in post-update value iteration introduces a novel framework for designing low-complexity MDP solutions based on domain-specific insights.

    \bibliographystyle{IEEEtran}
    \bibliography{ref}
\end{document}